\documentclass[prb, aps, twocolumn, amsmath, amssymb, superscriptaddress]{revtex4-2}
\usepackage{graphicx,verbatim}
\usepackage{braket}
\usepackage{color}

\usepackage{soul}
\usepackage[normalem]{ulem}
\usepackage[x11names]{xcolor}
\usepackage{hyperref}
\hypersetup{colorlinks=true,linkcolor=blue}

\newcommand{\tr}[1]{\text{Tr}\lbrace #1 \rbrace}

\begin{document}
\bibliographystyle{apsrev4-2}
\title{A Tensor Train Continuous Time Solver for Quantum Impurity Models}

\author{A. Erpenbeck}
\affiliation{
	Department of Physics, University of Michigan, Ann Arbor, Michigan 48109, USA
	}
\author{W.-T. Lin}
\affiliation{
	Department of Physics, University of Michigan, Ann Arbor, Michigan 48109, USA
	}
	
\author{T. Blommel}
\affiliation{
	Department of Physics, University of Michigan, Ann Arbor, Michigan 48109, USA
	}    
	
\author{L. Zhang}
\affiliation{
	Department of Physics, University of Michigan, Ann Arbor, Michigan 48109, USA
	}    

\author{S. Iskakov}
\affiliation{
	Department of Physics, University of Michigan, Ann Arbor, Michigan 48109, USA
	} 

\author{L. Bernheimer}
\affiliation{
	School of Chemistry, Tel Aviv University, Tel Aviv 6997801, Israel
	} 
 
\author{Y. N{\'u}{\~n}ez-Fern{\'a}ndez}
\affiliation{
	Universit\'e Grenoble Alpes, CEA, Grenoble INP, IRIG, Pheliqs, F-38000 Grenoble, France
	}    
\affiliation{
	Universit\'e Grenoble Alpes, CNRS, Institut N\'eel, F-38000 Grenoble, France
	}    

\author{G. Cohen}
\affiliation{
	School of Chemistry, Tel Aviv University, Tel Aviv 6997801, Israel
	}    
\affiliation{
	The Raymond and Beverley Sackler Center for Computational Molecular and Materials Science, Tel Aviv University, Tel Aviv 6997801, Israel
	}

\author{O. Parcollet}
\affiliation{
	Center for Computational Quantum Physics, Flatiron Institute, New York, New York, 10010, USA
	}    
 \affiliation{
	Universit\'{e} Paris-Saclay, CNRS, CEA, Institut de Physique Th\'{e}orique, 91191, Gif-sur-Yvette, France
	}    

\author{X. Waintal}
\affiliation{
	Universit\'e Grenoble Alpes, CEA, Grenoble INP, IRIG, Pheliqs, F-38000 Grenoble, France
	}    
	
\author{E. Gull}
\affiliation{
	Department of Physics, University of Michigan, Ann Arbor, Michigan 48109, USA
	}

\date{\today}

\begin{abstract}

The simulation of strongly correlated quantum impurity models is a significant challenge in modern condensed matter physics that has multiple important applications. Thus far, the most successful methods for approaching this challenge involve Monte Carlo techniques that accurately and reliably sample perturbative expansions to any order. However, the cost of obtaining high precision through these methods is high. Recently, tensor train decomposition techniques have been developed as an alternative to Monte Carlo integration. In this study, we apply these techniques to the single-impurity Anderson model at equilibrium by calculating the systematic expansion in power of the hybridization of the impurity with the bath. We demonstrate the performance of the method in a paradigmatic application, examining the first-order phase transition on the infinite dimensional Bethe lattice, which can be mapped to an impurity model through dynamical mean field theory. Our results indicate that using tensor train decomposition schemes allows the calculation of finite-temperature Green's functions and thermodynamic observables with unprecedented accuracy. The methodology holds promise for future applications to frustrated multi-orbital systems, using a combination of partially summed series with other techniques pioneered in diagrammatic and continuous-time quantum Monte Carlo.
\end{abstract}

\maketitle

\section{Introduction}
The solution of strongly correlated quantum impurity models is one of the central tasks of condensed matter physics. Quantum impurity models describe the physics of an interacting `impurity' or `quantum dot' coupled to a large, typically infinite, number of noninteracting `bath' or `lead' degrees of freedom.
Quantum impurity models were initially developed to describe the physics of magnetic impurities embedded in a non-magnetic host \cite{Anderson61}, but have since been adapted to describe quantum dots and molecular junctions \cite{Hanson07}, as well as atoms and molecules adsorbed on surfaces \cite{Brako81,Langreth91}. Importantly, they also appear as auxiliary models in quantum embedding techniques such as  dynamical mean field theory (DMFT) and self-energy embedding theory \cite{Georges92,Georges96,Kotliar06, Zgid17}, which typically require the calculation of a finite-temperature Green's function in the strongly correlated regime.

A reliable description of correlated systems requires methods that are numerically exact, in the sense that errors can be made arbitrarily small as a function of a control parameter. Among such methods, the continuous-time quantum Monte Carlo (CT-QMC) impurity solvers \cite{Rubtsov05,Werner06,Gull08,Gull11_RMP}, which are based on the stochastic sampling of a perturbative expansion to all orders, have become ubiquitous in cluster \cite{Maier05} and real-materials DMFT applications \cite{Kotliar06}. Numerous variants \cite{Werner06_Kondo,Haule07,Muhlbacher08,Gull10,Cohen15,Gunacker15,Eidelstein20,Li22}, improvements, and open source implementations \cite{Bauer11,Shinaoka14,TRIQS15, Seth16,Gaenko17,Shinaoka17,Yue19,Wallerberger19,Shinaoka20} exist.

CT-QMC methods provide exact results within Monte Carlo confidence intervals. In particular, they do not require a discretization of bath degrees of freedom, like exact diagonalization methods \cite{Caffarel_Exact_1994, Koch08,MejutoZaera20}, 
or of the time degrees of freedom, like \textcite{Hirsch86} or lattice Monte Carlo methods \cite{Blankenbecler81}.
However, they suffer from the following two limitations: (1) Away from high symmetry points, `sign problems' may cause the computational cost to grow exponentially as a function of system size, inverse temperature, or interaction strength, limiting calculations (with a few exceptions \cite{Eidelstein20,Li22}) in practice to systems with almost diagonal interactions and/or hybridizations, and (2) thermodynamic quantities such as the partition function and free energy are, in standard implementations, only available up to an unknown normalization constant \cite{Troyer03}. 

The standard framework of CT-QMC and, more generally, of `diagrammatic' \cite{Prokofev96,Prokofev98A,Prokofev98B} Monte Carlo methods is based on a perturbative series expansion that expresses observables of a quantum system in terms of an infinite series of high-dimensional integrals. This series is then summed to all orders in a stochastic sampling process, employing a Monte Carlo sampling procedure \cite{Gull11_RMP} that performs a random walk in diagram space. For a given number of samples $n_s$, this procedure produces unbiased stochastic estimates with errors that decrease rather slowly as $\sim 1/\sqrt{n_s}$. Notably, quasi-Monte Carlo methods can substantially accelerate this to $\sim 1/n_s$ in at least some cases~\cite{Macek20,Corentin21}.
In the context of correlated quantum transport, some of us \cite{NunezFernandez22} have recently shown that in the calculation of many high-dimensional integrals of the perturbative series expansion, it may be advantageous to replace Monte Carlo integration by a decomposition of the integrand into a product of low-dimensional tensors, which can then be integrated separately~\cite{dolgov_parallel_2020}.
The method is controlled in the sense that the exact result is recovered as the tensor rank increases.

The decomposition of tensors into approximate low-rank forms without needing to evaluate all their elements is based on tensor cross-interpolation (TCI) methods~\cite{goreinov_cross_2008,savostyanov2014}.  The corresponding approximation of a high-dimensional tensor by the product of a sequence of low-rank tensors is known as a `tensor train' in the applied mathematics and computer science literature~\cite{Oseledets2010}.
Tensor trains are also known as matrix product states (MPS) pioneered in the density matrix renormalization group (DMRG) and related methods~\cite{White92,Schollwoeck11}.

In this paper, we adapt the tensor train approach of \citet{NunezFernandez22}, to equilibrium imaginary-time quantum impurity problems. 
We benchmark the impurity solver for the analytically solvable noninteracting limit and demonstrate its accuracy for the single-site DMFT problem of a Bethe lattice in the infinite coordination number limit, where the DMFT is exact and the self-consistency condition becomes particularly simple \cite{Georges96}. 
We compute the Green's function (GF) to high accuracy, investigate convergence with respect to the parameters that control the accuracy, and showcase that the tensor train methodology is often substantially more precise than quantum Monte Carlo for a given amount of computer time.
In addition, we use the tensor train approach to compute the partition function and the impurity free energy. Unlike in CT-QMC, where normalization with respect to low order with quantum Wang--Landau \cite{Troyer03,Gull11_RMP} or normalization with respect to a hypervolume \cite{cohen_memory_2011,Cohen15} are needed, thermodynamic quantities are directly accessible in the tensor train decomposition scheme.
We then demonstrate the usefulness of our results at the example of the first-order Mott metal-to-insulator transition.

The paper proceeds as follows. Sec.~\ref{sec:method} introduces the hybridization expansion and gives an overview of the tensor train methodology, describing the decomposition for different observables, as well as computational details. Sec.~\ref{sec:results} illustrates results of the GF for the noninteracting and the DMFT case, and shows results for the free energy. Sec.~\ref{sec:conclusion} presents conclusions.

\section{Method}\label{sec:method}

This section introduces the methodology used in this work. The system and the hybridization expansion are presented in Sec.~\ref{sec:hyb}. The details of tensor train representation and its calculation are reviewed in Sec.~\ref{sec:tensor-train}. Details on how these two approaches are combined are presented in Sec.~\ref{sec:hyb+TT}.

\subsection{Hybridization expansion formalism}\label{sec:hyb}
We study a quantum impurity model described by the Hamiltonian $H = H_\text{I} + H_\text{B} + H_\text{IB} \equiv H_0 + H_\text{IB}$, consisting of an interacting impurity $H_\text{I}$, a noninteracting bath $H_\text{B}$, and the impurity--bath coupling or hybridization $H_\text{IB}$. 
For the single-site Anderson impurity model,
\begin{subequations}
\begin{align}
\label{eq:AM-I}
H_\text{I} &= \epsilon_0(n_\uparrow+n_\downarrow) + Un_\uparrow n_\downarrow ,\\
\label{eq:AM-B}
H_\text{B} &= \sum_{\sigma k} \epsilon_k c_{k\sigma}^\dagger c_{k\sigma},\\
\label{eq:AM-IB}
H_\text{IB} &= \sum_{k\sigma} \left(V_{k\sigma} d_\sigma^\dagger c_{k\sigma} + \text{h.c.}\right),
\end{align}
\end{subequations}
where $\epsilon_0$ denotes the on-site energy of the impurity and $U$ is the Coulomb interaction between two electrons of opposite spin. $k$ enumerates the (potentially infinite number of) bath states, $\epsilon_{k}$ is the dispersion of the noninteracting bath, and $V_{k\sigma}$ represents the coupling strength between the impurity and bath state $k$. 
The creation- and annihilation operators associated with spin-orbital $\sigma$ of the impurity are given by $d_\sigma^{(\dagger)}$, the operators $c_{k\sigma}^{(\dagger)}$ denote the corresponding bath operators associated with state $k$, while $n_\sigma = d_\sigma^{\dagger}d_\sigma$.

The main observables of interest for this work are the partition function, which grants access to thermodynamic properties of the system, and the Green's function (GF), which is of particular interest for quantum embedding schemes.
The partition function is given by 
\begin{align}
Z = \tr{e^{-\beta H}}, \label{Eq:Z}
\end{align}
where $\beta$ denotes the inverse temperature and $\tr{\dots}$ is the trace over the impurity and bath degrees of freedom.
We define the imaginary-time GF for electrons of spin $\sigma$ as
\begin{align}
G_{\sigma} (\tau) =  -\braket{T_\tau d_\sigma(\tau) d_{\sigma}^\dagger(0)} , \label{Eq:G}
\end{align}
where $\braket{\dots}$ denotes the expectation value with respect to the Hamiltonian $H$, $T_\tau$ is the time-ordering operator, and $d_\sigma(\tau) = e^{\tau H} d_\sigma e^{-\tau H}$.

In the hybridization expansion formalism, Eqs.~(\ref{Eq:Z}) and (\ref{Eq:G}) are expanded in orders of the impurity-bath coupling $H_\text{IB}$ \cite{Werner06}.
This hybridization expansion is one of the standard techniques underlying Monte Carlo quantum impurity solvers \cite{Gull11_RMP} and provides the framework for many approximate and numerically exact methods \cite{Keiter70,Pruschke89,Werner06,Werner06_Kondo,Haule07,Muhlbacher_Real_2008,Gull10_Bold,cohen_greens_2014,cohen_greens_2014-1,Cohen15,Eidelstein20}.

Expanding Eq.~(\ref{Eq:Z}) in the impurity--bath coupling yields
\begin{eqnarray}
    \label{Eq:Z_expan}
    Z &=&  \sum_{k=0}^{+\infty}
    \int_0^\beta \hspace{-0.2cm} d\tau_1 
        \int_{\tau_1}^\beta \hspace{-0.3cm} d\tau_2
        \cdots 
        \int_{\tau_{k-1}}^\beta \hspace{-0.6cm} d\tau_k\,
        \braket{ H_\text{IB}(\tau_k) \cdots H_\text{IB}(\tau_1) }_{H_0}  , \nonumber \\
\end{eqnarray}
with $H_0=H_\text{I}+H_\text{B}$,  $H_\text{IB}(\tau)=e^{H_0 \tau} H_\text{IB} e^{-H_0 \tau}$. $\braket{\dots}_{H_0}$ denotes the expectation value with respect to $H_0$.
Inserting the explicit expression for the impurity-bath coupling from Eq.~(\ref{eq:AM-IB}) and defining the time-ordered simplex $S_0^\beta$ as the region of integration with $0\leq\tau_1\leq\tau_2\leq\dots\leq\tau_k\leq\beta$, the partition function can be reexpressed as
\begin{eqnarray}
    Z &=& \sum_{k=0}^\infty 
            \int_{S_0^\beta} d\tau_1 \cdots d\tau_k\,
            \sum_{\sigma_1 \dots \sigma_k}
            z^{(k)}_{\sigma_1 \dots \sigma_k}(\tau_1,\dots,\tau_k)\,. \nonumber \\ \label{eq:integral_expression_Z}
\end{eqnarray}
with
\begin{eqnarray}
    z^{(k)}_{\sigma_1 \dots \sigma_k}(\tau_1,\dots,\tau_k)
        &=&
            \hspace{-0.15cm}
            \sum_{\phi_1 \dots \phi_k}
            \hspace{-0.15cm}
            \det\mathbf{\Delta} 
            \braket{ d_{\sigma_k}^{\phi_k}(\tau_k) \cdots d_{\sigma_1}^{\phi_1}(\tau_1)}_{H_{\text{I}}}
            .
            \nonumber \\
            \label{eq:zk}
\end{eqnarray}
Here, $\sigma_i\in\lbrace\uparrow, \downarrow\rbrace$ are spin-indices and $\phi_i\in\lbrace -, +\rbrace$ are used to sum over all combinations of creation and annihilation operators with $d_{\sigma_i}^- \equiv d_{\sigma_i}$ and $d_{\sigma_i}^+ \equiv d_{\sigma_i}^\dagger$.
The influence of the bath on the impurity is encoded in the hybridization function $\Delta(\tau)$.
The hybridization function can either be calculated explicitly for a given bath model---which is done for studies on quantum dots or molecular systems
where $\Delta(\tau) = -\braket{T_\tau a_\sigma(\tau)a_\sigma^\dagger(0)}$  with $a_\sigma = \sum_k V_{k\sigma} c_{k\sigma}$
---or determined by a self-consistency condition.
The latter scenario appears in quantum embedding schemes like  DMFT.
Given a hybridization function $\Delta(\tau)$ and invoking particle-hole symmetry, the hybridization matrix $\mathbf{\Delta}$ entering Eq.~(\ref{eq:zk}) is
\begin{eqnarray}
    \mathbf{\Delta}_{ij} &=& \begin{cases}
                                \Delta(\tau_i-\tau_j)\quad &\text{if }\sigma_i=\sigma_j\text{ and }\phi_i\neq\phi_j,\\
                                0 &\text{otherwise} .\\
                            \end{cases}
\end{eqnarray}

The hybridization expansion expressions for the GF can be obtained in a similar fashion \cite{Werner06, Gull11_RMP}, and assume the form
\begin{equation}
    G_{\sigma}(\tau) = \sum_{k=0}^\infty 
        \int_{S_0^\beta} 
        \hspace{-0.2cm}
        d\tau_1 \dots  d\tau_k
        \hspace{-0.2cm}
        \sum_{\sigma_1 \dots \sigma_k}
        g^{(k)}_{\sigma\sigma_1 \dots \sigma_k}(\tau, \tau_1,\tau_2,\dots,\tau_k)\,.\label{eq:integral_expression_GF}
\end{equation}
Here,
\begin{eqnarray}
    \label{eq:gk}
    g^{(k)}_{\sigma\sigma_1 \dots \sigma_k}(\tau, \tau_1,\dots,\tau_k)
        &=& - \frac{1}{Z} \sum_{\phi_1 \dots \phi_k}   
            \det\mathbf{\Delta} \times \\ &&
            \hspace{-3.25cm}
             \braket{ d_{\sigma_k}^{\phi_k}(\tau_k) \cdots 
            d_{\sigma_m}^{\phi_m}(\tau_m) d_\sigma(\tau) d_{\sigma_n}^{\phi_n}(\tau_n)
            \cdots d_{\sigma_1}^{\phi_1}(\tau_1)d_\sigma^\dagger}_{H_{\text{I}}} 
            \nonumber        
\end{eqnarray}
and $\tau_m \geq \tau \geq \tau_n$.
While the main difference between the integrands $z^{(k)}_{\sigma_1 \dots \sigma_k}$ and $g^{(k)}_{\sigma\sigma_1 \dots \sigma_k}$ are the creation and annihilation operators at $0$ and $\tau$, the additional operators further restrict the combinations of spin and creation and annihilation operators that give a nonzero contribution. 

Eq.~(\ref{eq:zk}) and (\ref{eq:gk}) contain sums over all possible creation and annihilation operator combinations $\phi_i$, which  leads to an exponential number of possible combinations of operators. In the special case of density-density interactions,  there is only a single non-zero contribution with alternating creation and annihilation operators for time-ordered arguments. This simplification in the density-density case is analogous to the simplification to the `segment' picture in CT-HYB \cite{Werner06,Gull11_RMP}.

The hybridization expansion presented here represents a \textit{bare} expansion scheme. A variety of related partial summation schemes have also been  successful~\cite{Prokofev08,vanHoucke12,Gull10_Bold,cohen_greens_2014,Cohen15,Eidelstein20}.

Eqs.~(\ref{eq:integral_expression_Z}) and (\ref{eq:integral_expression_GF}) describe an infinite series of terms in a series expansion, where a contribution at order $k$ consists of a $k$-dimensional integral.
For finite systems at finite temperature, this series is convergent \cite{Werner06,Rubtsov05}.
However, since the largest contributions to the series typically comes from orders near $\beta \langle H_\text{IB}\rangle$ \cite{Haule07}, contributions at increasingly high orders are expected when the temperature is lowered.
Traditionally, the expressions in Eqs.~(\ref{eq:integral_expression_Z}) and (\ref{eq:integral_expression_GF}) are evaluated by Monte Carlo techniques, 
whereby the integrands 
$z^{(k)}_{\sigma_1 \dots \sigma_k}$ and $g^{(k)}_{\sigma\sigma_1 \dots \sigma_k}$ can be interpreted in terms of Feynman diagrams, which are then combined in a determinant and summed over in a statistical manner~\cite{Werner06}.

\subsection{Tensor Train Decomposition and TCI}\label{sec:tensor-train}

\begin{figure}[tb]
\raggedright a)\\
\centering
\includegraphics[width=0.45\textwidth]{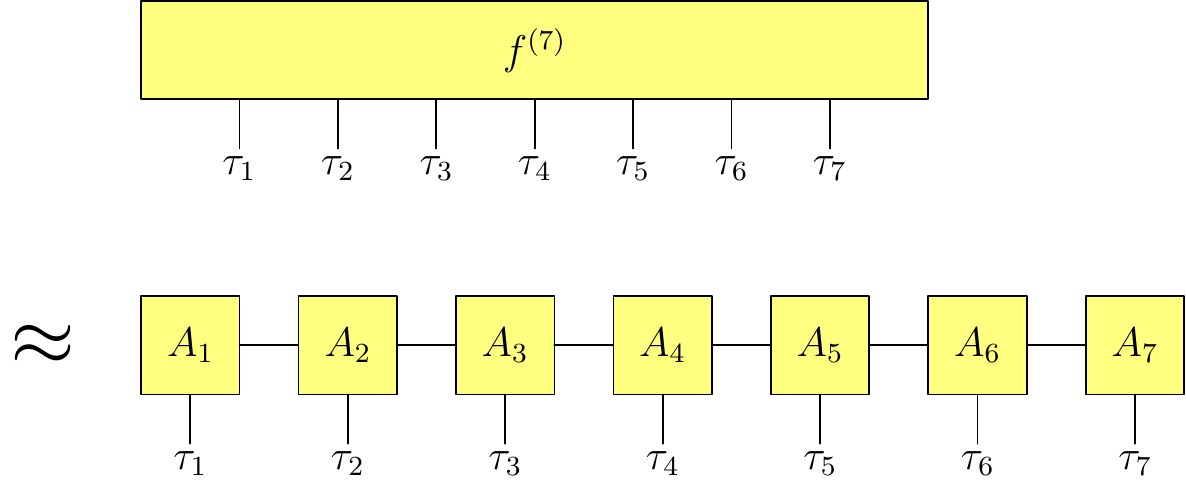}\\
\raggedright b)\\
\centering
\includegraphics[width=0.3\textwidth]{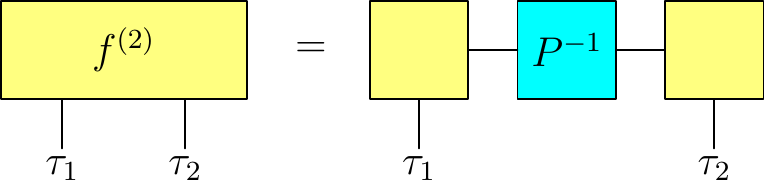}\\
\raggedright c)\\
\centering
\includegraphics[width=0.4\textwidth]{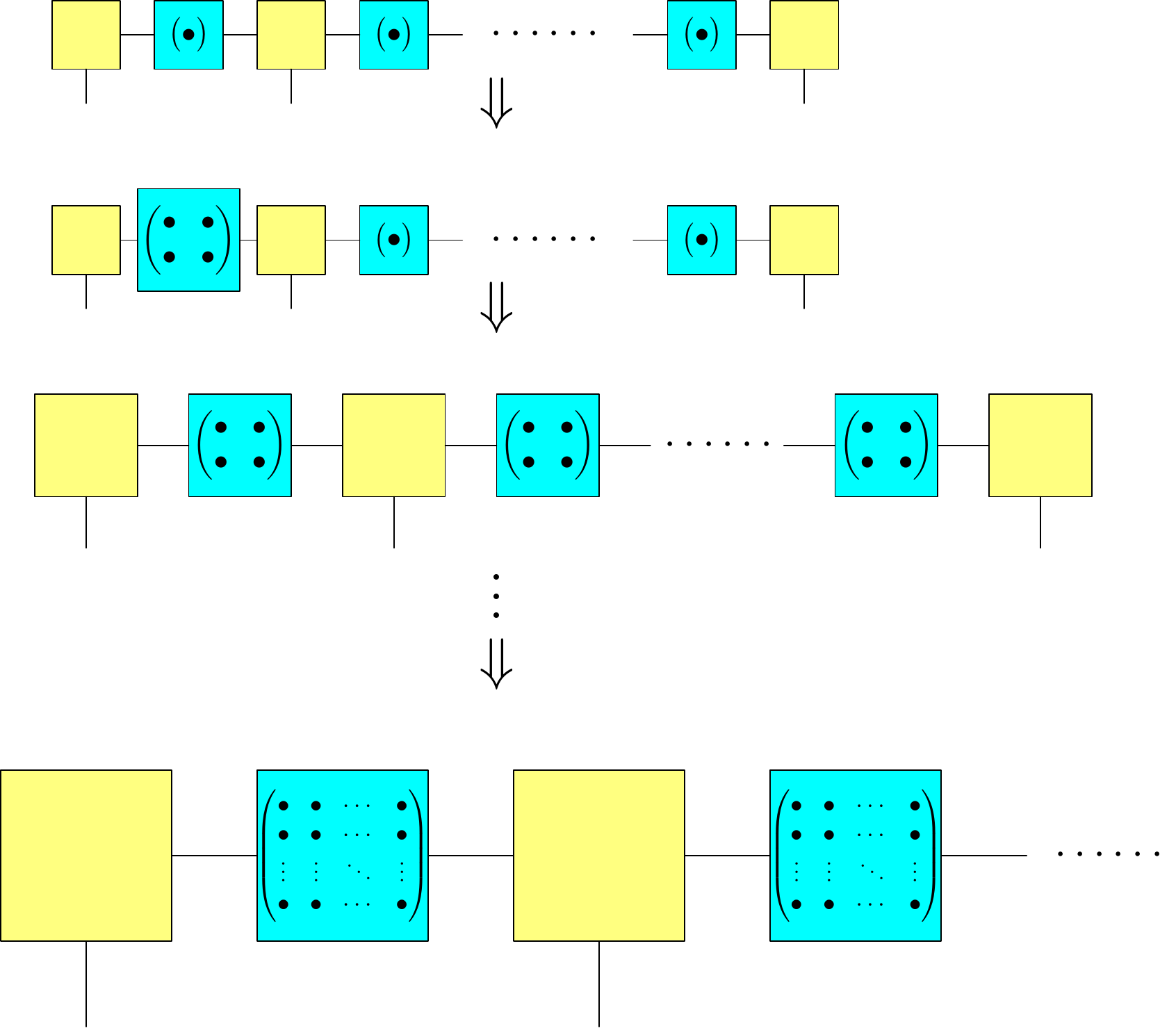}\\
\caption{
        a: Visualization of the tensor train representation of a function $f^{(k)}\left(\tau_{1},\tau_{2},\dots,\tau_{k}\right)$ according to Eq.~(\ref{eq:tt-decomp}) for $k=7$. Each block labeled $A_i$ with $1\leq i \leq k$ is either a matrix valued function of $\tau_i$, or a tensor in case that the $\tau_i$ are discretized on a grid. Horizontal lines between blocks imply matrix multiplication, vertical lines denote the dependence on $\tau_i$.
        b: Depiction of Eq.~(\ref{eq:2d-tt}), where the pivot matrix is used to decompose a two-dimensional function into a tensor train.
        c: Visualization of the iterative scheme used to approximate a multi-dimensional integrand by a tensor train.
        }
        \label{fig:tensor-train}
\end{figure}

In order to calculate observables such as the partition function or the GF within the hybridization expansion, the high-dimensional integrals in Eqs.~(\ref{eq:integral_expression_Z}) and (\ref{eq:integral_expression_GF}) need to be evaluated. Traditionally, these integrals are calculated using Monte Carlo techniques \cite{Rubtsov05, Werner06}, which converge as $\sim 1/\sqrt{n_s}$ for a given number of stochastic samples $n_s$.
The tensor train methodology offers an alternative approach that has potential to converge faster than Monte Carlo. In the following, we summarize the principles underlying  the tensor train representation and refer the reader to the applied mathematics literature for mathematical proofs   \cite{Oseledets2010,savostyanov2014,dolgov_parallel_2020}. The implementation of the tensor-fitting component of this work follows the paper of \citet{NunezFernandez22}.

To motivate the construction of a tensor train representation for a given integrand, we first consider the general task of integrating a high-dimensional function $f(\tau_1,\tau_2,\dots,\tau_k)$ over all its coordinates,
\begin{equation}
I=\int_{\boldsymbol{\tau} \in [0,\beta]^{k}}\hspace{-1cm}\mathrm{d}\tau_{1}\mathrm{d}\tau_{2}\cdots\mathrm{d}\tau_{k} \ f\left(\tau_{1},\tau_{2},\dots,\tau_{k}\right), \label{eq:int_full}
\end{equation}
which is similar (though not yet equivalent) to evaluating Eqs.~(\ref{eq:integral_expression_Z}) and (\ref{eq:integral_expression_GF}).
If the integration variables are separable and independent, i.e. $f(\tau_1,\tau_2,\dots,\tau_k)=f_1(\tau_1)\cdots f_k(\tau_k)$, the integral can be reexpressed as $k$ one-dimensional integrals,  
\begin{equation}
\begin{aligned}
 I 
 & =\left[\int_0^\beta\mathrm{d}\tau_{1}f_1\left(\tau_{1}\right)\right]\cdots\left[\int_0^\beta\mathrm{d}\tau_{k}f_k\left(\tau_{k}\right)\right],
 \label{eq:int_factor}
\end{aligned}
\end{equation}
which can be evaluated independently with standard quadrature rules.
Assuming that the arguments $\tau_i$, with $1\leq i\leq k$, are each represented on a quadrature grid with $n_\tau$ points, the complexity of evaluating Eq.~(\ref{eq:int_full}) is reduced from $n_\tau^k$ to $k n_\tau$ for Eq.~(\ref{eq:int_factor}) -- a substantial improvement, especially for large values of $k$.

The tensor train representation can be viewed as a generalization of this special case.
While an arbitrary function $f^{(k)}\left(\tau_{1},\tau_{2},\dots,\tau_{k}\right)$ might not be separable in its arguments, we aim to construct an approximation
\begin{equation}
\label{eq:tt-decomp}
  f^{(k)}\left(\tau_{1},\tau_{2},\dots,\tau_{k}\right) \simeq A_1(\tau_1)A_2(\tau_2)\cdots A_k(\tau_k),  
\end{equation}
where the $A_i(\tau_i)$ are matrices of dimension $r_{i-1}\times r_i$ for $1<i<l$ with $r_1=r_{k}=1$. Matrix multiplication between the $A_i(\tau_i)$ is implicit in this notation. 
This representation is schematically visualized in Fig.~\ref{fig:tensor-train}a.
As the decomposition in Eq.~(\ref{eq:tt-decomp}) has the same temporal structure as Eq.~(\ref{eq:int_factor}), it allows for the same simplification when evaluating the integral, the only difference being that the components are matrix valued functions of $\tau_i$.
In the discretized case where the $\tau_i$ are on a grid, $A_i(\tau_i)$ can be interpreted as a tensor rather than a matrix valued function, for which the above statements also hold. We refer to a representation of the form in Eq.~(\ref{eq:tt-decomp}) as a tensor train representation.

A tensor train representation (or approximation) is said to be of rank $r$ if the matrices (or tensors in the discretized case) in Eq.~(\ref{eq:tt-decomp}) are of dimension $r\times r$ (or $r\times n_\tau\times r$).
The construction of low-rank tensor representations has been studied extensively~\cite{goreinov_cross_2008,Oseledets2010,oseledets2013,savostyanov2014,khoromskij2018tensor}.
Here, we use an extension by \citet{dolgov_parallel_2020} of the algorithmic ideas of \citet{Oseledets2010}, which are based on the so-called cross interpolation~\cite{Tyrtyshnikov2000}, and which is applicable to matrices and tensors.
The basic idea is that a set of points $\left(\tau_{1},\tau_{2},\dots,\tau_{k}\right)$, which are called pivots, defines an approximation for the function $f^{(k)}$ at all possible values of the times.
The approximation requires evaluating the function at the pivots themselves; and at all possible values of a certain coordinate, with other coordinates held constant.
Essentially, evaluation is performed on sets of 1D lines in the high-dimensional hypercube of all coordinates, which cross through pivots.

To gain some intuition, it is useful to first consider the special case of matrices, i.e. $k=2$. Let us assume that one knows the values in a matrix only at a certain subset of its rows and columns, defined by a set of pivot coordinates where they cross. It is possible to obtain an interpolation scheme based on this partial information.
For a function $f^{(2)}\left(\tau_{1},\tau_{2}\right)$,
evaluating $f^{(2)}$ at the set of pivots $\tau_{1j}$ and $\tau_{2j}$ with $1\leq j\leq r$ with the total number of pivots $r$, one can construct the pivot matrix $P$ whose entries are given by $P_{j j'} = f^{(2)}\left(\tau_{1j},\tau_{2j'}\right)$. Using the inverse of the pivot matrix, one obtains the cross interpolation of the original function,
\begin{align}
    f^{(2)}\left(\tau_{1},\tau_{2}\right)\!\simeq\!\!\sum_{j, j' =1}^r\!\!f^{(2)}\left(\tau_{1},\tau_{2j}\right) [P^{-1}]_{jj'} f^{(2)}\left(\tau_{1j'},\tau_{2}\right) ,\label{eq:2d-tt}
\end{align}
which can be visualized in tensor network form as in Fig.~\ref{fig:tensor-train}b.
Eq.~(\ref{eq:2d-tt}) represents an interpolation of $f^{(2)}\left(\tau_{1},\tau_{2}\right)$ in the sense that it is exact if $\tau_1$ (or $\tau_2$) belong to the set
$\tau_{1j}$ (or $\tau_{2j}$). 
Moreover, if $f^{(2)}$ is of rank $r$, {\it i.e.} it can be expressed as $f^{(2)}\left(\tau_{1},\tau_{2}\right) = \sum_{j,j'=1}^r f_{1j}(\tau_1) f_{2j'}(\tau_2)$ then Eq.~(\ref{eq:2d-tt})  becomes exact when one uses $r$ pivots
provided that  $P$ remains invertible. 
Generally, a given approximation of this type can be systematically improved by sequentially introducing more pivots into it.
However, this rapidly becomes costly, and not all new pivots provide the same amount of information.
While an optimal procedure remains unknown, there are well-established heuristic algorithms for systematically finding and incorporating pivots into the approximation in such a way that convergence occurs rapidly \cite{goreinov_pseudo_1997, Goreinov_Matrix_2010}. 

After establishing the TCI for a function of two variables, we outline the extension to functions with multiple discrete variables $f^{(k)}\left(\tau_{1},\tau_{2},\dots,\tau_{k}\right)$.
The approximation is initialized by considering a single pivot, so that we have pivot matrices of size one (see top row of Fig. 1c).
We then perform a search along the first two dimensions, $\tau_1$ and $\tau_2$, for the next pivot to be added; all other coordinates are held constant at the value of the original pivot.
The chosen pivot is used to enlarge the leftmost pivot matrix to $2\times 2$ (2nd row of Fig. 1c).
In a manner reminiscent of the density matrix renormalization group algorithm, we subsequently sweep to the right, repeating this procedure by starting from existing pivots and modifying coordinates locally, until all pivot matrices are of rank $2$ (3rd row of Fig. 1c).
This is followed by another sweep to the left, resulting in rank $3$ pivot matrices.
This procedure repeats until the pivot matrices have reached an initially specified maximum rank (Fig. 1c, bottom column) \cite{Oseledets2010,savostyanov2014}.

The search for suitable pivots and the addition of new pivots is an important component in the algorithm, for which we follow the `maximum volume' procedure described by \textcite{dolgov_parallel_2020}.
At each step, we consider the 4D tensor comprising the product of a single (cyan inf Fig.~\ref{fig:tensor-train}) pivot matrix and its two adjoining (yellow) tensors.
We search this 4D space for candidate pivots by evaluating the function on a series of its 2D subspace. The objective of the search is to find a pivot maximizing the approximation error.
Because the tensor train approximation is an interpolation, this maximal error decreases to zero once this pivot is included in the tensor train.

For multi-dimensional continuous variables $\{\tau_1,\dots,\tau_k\}$, we choose a set of collocation points for each $\tau_i$, such that the multivariate function is discretized into a multi-dimensional tensor and the discrete algorithm above can be applied.

\subsection{Applying the tensor train approximation to expressions from the hybridization expansion}\label{sec:hyb+TT}

The efficiency of a tensor train approximation for a given integration task depends on whether an accurate approximation for an integrand can be found for low rank $r$. 
In the following, we describe some of the technical aspects of our implementation of the tensor decomposition scheme to the hybridization expansion.
We outline the main aspects for the partition function, and then discuss the specific aspects that are needed to apply the method to the GF.

\subsubsection{Mapping the hypercube to the simplex}
The expressions from the hybridization expansion framework require the integration over the simplex $S_0^\beta$
see Eqs.~(\ref{eq:integral_expression_Z}). 
This is a consequence of the time-ordering.
However, the TCI algorithm is defined on the hypercube.
Simply extending the integral beyond the time-ordered region would introduce discontinuities that prevent an accurate low-rank tensor train approximation. 

The change of variable proposed in Ref.~\cite{NunezFernandez22}, when applied to imaginary-time problems, extends the hybridization function $\Delta(\tau)$ over the discontinuities at $\tau=0$ and $\tau=\beta$ and is therefore also not suitable.

Instead, we use a change of variable in this work that maps the original simplex $S_0^\beta$ to the
hypercube $[0, 1]^k$.

Let $\tau_i$ denote the  variables within the simplex and $v_i$ the corresponding variables in the hypercube with $1\leq i \leq k$. The mapping $h$ between the hypercube and the simplex used in this work is
    \begin{subequations}
    \label{eq:simplex}
    \begin{eqnarray}
        \tau_1 &= h(v_1) &= \chi(v_1)\cdot \beta ,  \label{eq:simplex1}\\
        \tau_j &= h(v_j) &= \tau_{j-1} + \chi(v_j)\cdot (\beta - \tau_{j-1}) \label{eq:simplex2},
    \end{eqnarray}
    \end{subequations}
with $2\leq j \leq k$ and $\chi$ any differentiable monotonous function that maps the interval $[0,1]$ onto itself. A simple choice of $\chi$ is the identity $\chi(x)=x$ but a different choice of $\chi$, such as $\chi(x)=x^2$, may facilitate
the TCI.

\begin{figure}[tb]
    \centering
    \includegraphics[width=0.45\textwidth]{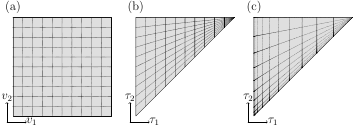}
    \caption{
            Illustration of the mapping between the hypercube (panel a) and the simplex in two dimensions, using $\chi(x)=x$ (panel b) and $\chi(x)=x^2$ (panel c). Each intersection of lines in panel a is mapped into an intersection in panel b and c.
            }
            \label{fig:change_of_var}
    \end{figure}

The two different mappings are visualized in Fig.~\ref{fig:change_of_var}. Panel a shows a uniform partitioning of the hypercube, which is mapped to the simplex by $\chi(x)=x$ in panel b, and by $\chi(x)=x^2$ in panel c. 

\subsubsection{Change of variables}
 To proceed, we calculate the Jacobian of the change of variable
 defined by Eqs.~(\ref{eq:simplex1}) and (\ref{eq:simplex2}). The Jacobian matrix $J_{ij} = \partial\tau_i/\partial v_j$ is upper triangular so that its determinant is given by the product of its diagonal elements,

    \begin{eqnarray}
        \det J(v_1, \dots, v_k) &=& \beta \chi'(v_1) \cdot \prod_{j=2}^{k} \chi'(v_2) (\beta-h(v_j)). \nonumber \\ \label{eq:Jabobian}
    \end{eqnarray}
While it is in principle possible to apply TCI separately to the Jacobian and the integrand
$\tilde z^{(k)}_{\sigma_1 \dots \sigma_k}(v_1,\dots,v_k)$, we choose to apply TCI directly on their product, ie.\ we apply the tensor train approximation to
\begin{eqnarray}
        \bar z^{(k)}_{\sigma_1 \dots \sigma_k}(v_1,\dots,v_k) &=& \tilde z^{(k)}_{\sigma_1 \dots \sigma_k}(v_1, \dots,v_k)  
        \lvert\det J(v_1, \dots, v_k)\rvert . \nonumber \\\label{eq:zk+det} 
\end{eqnarray}
This approach proved to be the most efficient within the scope of this work, since the factorization of Eq.~(\ref{eq:zk+det}) by a tensor train allows for a direct calculation of the integral using one-dimensional quadrature rules.
    
    The function $\chi$ that enters Eqs.~(\ref{eq:zk+det}) through Eqs.~(\ref{eq:simplex1}), (\ref{eq:simplex2}), and (\ref{eq:Jabobian}) controls two important aspects that influence the quality of low-rank tensor train approximations. 
    First, it influences the spacing of the pivot points that are used within the tensor train decomposition. For example, the identity for $\chi$ in Eqs.~(\ref{eq:simplex1}) and (\ref{eq:simplex2}) shifts potential pivot points away from $0$ and closer to $\beta$, which implies a bunching of pivot points close to $\beta$.  This can be compensated for by a suitable choice of $\chi$.
    Second, as $\chi$ enters the integrands, it can be used to `warp' the integrand i.e. reshape it into a function that is easier to integrate. The approach of `warping' the integrand was also used for facilitating integration using Quasi-Monte Carlo methods \cite{Macek20}, and was used in Ref.~\cite{NunezFernandez22}, where changing to a representation in terms of relative time arguments rather than absolute ones was necessary to construct a tensor train approximation.
    For the results presented in this work, we use $\chi(v_i) = v_i^2$. 
    
\subsubsection{Summation over spin indices:}
Eq.~(\ref{eq:integral_expression_Z}) features a sum over spin indices $\sigma_i$.
The number of spin combinations that need to be considered grows exponentially with the hybridization order $k$, which can render the calculation of higher orders in the hybridization expansion unfeasible. We have considered two different approaches for incorporating this aspect into the tensor train methodology.
    
    The first approach is to use the sum over all spin combinations for the tensor train decomposition explicitly, that is applying the TCI algorithm on $\sum_{\sigma_1 \dots \sigma_k} \bar z^{(k)}_{\sigma_1 \dots \sigma_k}(v_1,\dots,v_k)$.
    As this approach effectively averages over all spin combinations, which smooths to some extent the function that is approximated, we found that this method produces good approximations for relatively low tensor ranks. However, as the evaluation of the function that is approximated by the tensor train requires to explicitly perform the sum over all spin combinations which grows exponentially with the hybridization order, this approach becomes prohibitively expensive for high hybridization orders. As such, we deem this approach only feasible for high temperature and for systems that converge within hybridization orders of $k\lesssim 15$. 

    The second approach is to use the tensor train approximation not only for the arguments $v_i$, but also for the spin arguments $\sigma_i$. The sum over the spin indices is then performed along with the integration over the simplex.
    On a technical level, we achieve this by introducing a surrogate variable $w_i\in[-1, 1]$ with $v_i=|w_i|$ and $\sigma_i=\uparrow$ if $w_i\geq0$ and $v_i=|w_i|$ and $\sigma_i=\downarrow$ if $w_i<0$.
    Approximating $\bar z^{(k)}(w_1,\dots,w_k)$ by a tensor train and integrating over the variables $w_i$ corresponds to summing over all spin indices and integrating over the original simplex.
    This approach overcomes the necessity to explicitly account for an exponential number of spin combinations and is therefore suitable for high hybridization orders. However, as the function that is being approximated in this case contains more information, the tensor rank that is required to obtain a certain accuracy increases as compared to the previous approach.

\subsubsection{Decomposition schemes for the GF}
    The $\tau$-dependence of the GF is the result of the operator $d_\sigma$ placed at time $\tau$, which is the main difference between the partition function and the GF. 
    When performing the $\tau_i$-integrals in Eq.~(\ref{eq:integral_expression_GF}), this implies that the integrand $g^{(k)}_{\sigma\sigma_1 \dots \sigma_k}(\tau, \tau_1,\dots,\tau_k)$ needs to consider configurations that have a variable number of up to $k$ creation and annihilation operators to the left or to the right of $\tau$.
    When creation or annihilation operators move across $\tau$, where the operator $d_\sigma(\tau)$ is located, the integrand drops to zero and a discontinuity occurs. Tensor train approximations converge slowly in the presence of such discontinuities.
    We therefore rewrite the integration over the simplex in Eq.~(\ref{eq:integral_expression_GF}) as
    \begin{eqnarray}
        G_{\sigma}(\tau) &= &
        \sum_{k=0}^\infty
        \sum_{l=0}^k
        \int_{S_0^\tau}
            d\tau_1 \dots  d\tau_l
        \int_{S_\tau^\beta}
            d\tau_{l+1} \dots  d\tau_k
        \nonumber \\ &&
        \sum_{\sigma_1 \dots \sigma_k}
        g^{(k)}_{\sigma\sigma_1 \dots \sigma_k}(\tau, \tau_1,\tau_2,\dots,\tau_k)\, , \label{eq:split_GF}
    \end{eqnarray}
    which fixes the number of creation and annihilation operators to the right and to the left of $\tau$ to $l$ and $k-l$, respectively, thus circumventing the emergence of discontinuities. 
    Here, $S_0^\tau$ and $S_\tau^\beta$ are the time-ordered simplices between $0$ and $\tau$ and between $\tau$ and $\beta$, respectively, with $0\leq\tau_1\leq\dots\leq\tau_l\leq\tau$ and $\tau\leq\tau_{l+1}\leq\dots\leq\tau_k\leq\beta$.
    
To represent the $\tau$-dependence of the GF, we have explored two possible approaches.
In the first one, we calculate the GF at every value of $\tau$. This approach scales linearly with the number $\tau$-points.

In the second one, we employ the tensor train approach to also interpolate the $\tau$ dependence. On a technical level, this is done by adding $\tau$ as a parameter to the function that is approximated by a tensor train. Generally, this requires a higher rank tensor approximation for comparable accuracy, whereby the actual increase in numerical effort depends on the details of the problem and the representation used for $\tau$. We chose the second method for the calculations reported below, where we find that we need about twice the rank to obtain results of comparable accuracy.

\subsubsection{Numerical stability and current limitations}
While the tensor train methodology is generally a powerful tool to obtain highly accurate results, we encountered cases where our implementation of the approach became unstable.
        
Unsurprisingly, we observed that the tensor train method fails to provide accurate results whenever it is applied to functions that are discontinuous. The issue can be avoided by ensuring that this case is not encountered for the observable of interest.
    
Second, we observed that the tensor train approach described in this work may become unstable for high hybridization orders, $k\gtrsim30$, while at the same time assessing convergence for these high orders  becomes challenging. These issues are likely caused by the selection of pivots. As only a fraction of the vast parameter space can be probed for high dimensional functions, pivots might be chosen in such a way that they do not provide a good representation of the function that is being interpolated. Moreover, the CI scheme relies on the inversion of the pivot matrix. At high orders, where the function approximated is essentially zero for extended regimes of the parameter space, picking near-singular pivots for the CI scheme may result in an imprecise tensor train approximations. Further investigation of these numerical aspects will likely resolve the issue.

\section{Results}\label{sec:results}
In this section, we present results for the single-impurity Anderson impurity model as described by Eqs.~(\ref{eq:AM-I})--(\ref{eq:AM-IB}). 
The influence of the bath on the impurity is encoded in the hybridization function $\Delta(\tau)$.

We benchmark our method for the exactly solvable case of a noninteracting impurity in Sec.~\ref{sec:nonint}, which we use to assess the accuracy that can be obtained by the present method. In Sec.~\ref{sec:DMFT}, we showcase the performance of the methodology for the paradigmatic metal-to-insulator transition observed in the infinite dimensional Bethe lattice. In particular, we show that the method not only provides accurate and noise-free results for the GF, but grants direct access to thermodynamic properties. We illustrate this  with the metal-to-insulator transition.

\subsection{Noninteracting limit} \label{sec:nonint}
We showcase the performance our method for the case of a noninteracting Anderson impurity model, $U = \epsilon_0 = 0$, which is also known as the resonant level model.
The system is analytically solvable, see e.g. \cite{Bruus_Many_2004, Haug_Qauntum_2008}, which allows us to benchmark the precision of the results obtained from a tensor train approximations of different ranks.
Nevertheless, it is a challenging benchmark for hybridization expansions approaches as it performs an expansion around the `atomic' limit of an isolated impurity \cite{Werner06}.

We consider an impurity coupled to a bath whose dispersion $\epsilon_k$ has a semi-elliptical form with bandwidth $4t$; the associated density of states is 
$D(\omega) = \frac{1}{\pi t^2} \sqrt{4 t^2 - \omega^2}$ for $-2t\leq\omega\leq2t$ and the hybridization function is given by $\Delta(\tau) = -\int d\omega D(\omega) \frac{e^{-\tau\omega}}{1+e^{-\beta\omega}}$. 
This system presents a paradigmatic case studied in single-site DMFT, as it corresponds to an impurity embedded in an infinite-dimensional Bethe lattice \cite{Georges96}.

\begin{figure}[tb]
            \raggedright a)\\
            \centering     \includegraphics{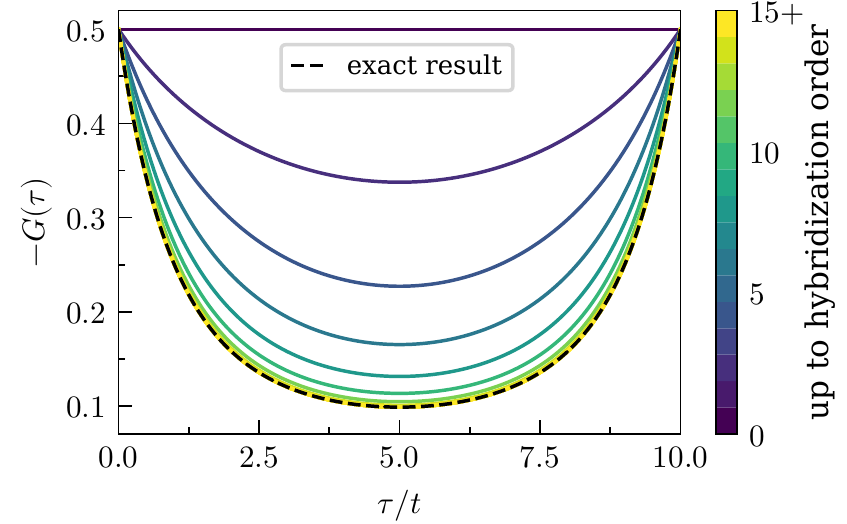}\\
            \raggedright b)\\
            \includegraphics{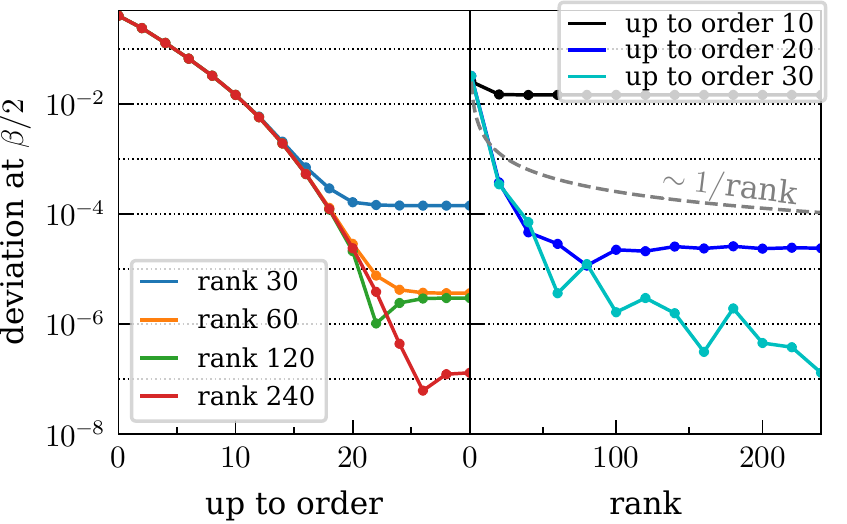}
            \caption{
                Noninteracting impurity coupled to a semi-elliptic bath with bandwidth of $4t$ at temperature $\beta=10/t$.
                a: Contribution to GF from terms up to hybridization orders indicated by the color. 
                b: Deviation of $G(\beta/2)$ from the analytic result as a function of perturbative orders for a set of representative tensor ranks (left panel), and as a function of tensor rank for a set for representative perturbative orders (right panel).}
            \label{fig:GF_nonint}
\end{figure}
Fig.~\ref{fig:GF_nonint}a shows the convergence of the GF of the tensor train formalism described in this work to the exact result as a function of expansion order at inverse temperature $\beta =10/t$.
'Exact' denotes the analytically known result. 
As is evident, the exact result is recovered (within the accuracy of this plot) as the expansion is increased beyond an hybridization expansion order $k\sim 10.$

In order to further assess the accuracy, we consider the deviation of the GF at $\beta/2$, {\it i.e.} in the middle of the interval.
The left panel of Fig.~\ref{fig:GF_nonint}b shows this deviation, as a function of expansion order, for different tensor decomposition ranks. 
We see that a maximum precision of $10^{-4}$ can be reached  for a decomposition rank of $30$. Adding contributions at higher order does not make the result more precise, indicating that it is the tensor rank, rather than the truncation of the expansion at a given order, that limits this precision. 
This is corroborated by the curves for rank $60$, rank $120$ and rank $240$, which systematically increase the precision of the GF to an accuracy of $10^{-7}$. 
Higher accuracy is reached by a combination of increasing diagram order and increasing tensor rank; higher order hybridization contributions require higher ranks to be accurately approximated by a decomposition.

The right panel of Fig.~\ref{fig:GF_nonint}b illustrates the same behavior as a function of tensor rank, evaluating contributions at up to hybridization order $10$, $20$, and $30$.
The grey dashed line indicates a convergence $\sim1$/rank with respect to the tensor rank. In particular, it shows that the method converges faster than $1/$rank for higher orders.
It is evident that while contributions up to order $10$ are well described by an approximation tensor trains of rank less than $50$, higher order contributions require substantially higher tensor ranks.

In practice, these results suggest a scheme where tensor train approximations for a fixed hybridization are performed for gradually increasing tensor ranks, until the integral values no longer change as a function of tensor rank. 
Note that it is difficult with existing CT-QMC techniques to reach a relative accuracy beyond $10^{-5}$; the tensor train methodology is therefore promising for obtaining high-precision data that could be used, for example, in analytic continuation \cite{Fei21}.

\subsection{Dynamical Mean-Field Theory and Free Energy} \label{sec:DMFT}
In the following, we present results for interacting impurity models.
The purpose of this section is two fold: First, we demonstrate the performance of our method for interacting system and its capability to generate results that are compatible with findings that are obtained using CT-QMC.
Second, we show that our method grants direct access to thermodynamic observables, which are not straightforwardly available in standard CT-QMC methods. 

For scope of this section, we study the paradigmatic example of the metal-to-insulator transition in an infinite dimensional Bethe lattice as described within DMFT \cite{Georges92, Jarrell_Hubbard_1992, Rozenberg_Mott_1992, Zhang_Mott_1993, Rozenberg99, Bulla99,  Bluemer_PhD}.
In its single-site formulation, DMFT approximates the momentum-dependent self-energy of an extended lattice problem by a local self-energy, and then provides a solution for the auxiliary impurity problem with a dynamically adjusted, self-consistently determined bath \cite{Georges92,Georges96}. 
For the infinite coordination-number Bethe-lattice, the self-energy is local and the methodology becomes exact \cite{Metzner_Correlated_1989, Georges92, Georges96, Eckstein_Hopping_2005}. 
Below a critical temperature of $\beta \sim 15/t$, the paramagnetic version of the model is known to have a first-order Mott metal-to-insulator transition between a metallic state at weak interaction and an insulating state at large interaction, with an extended coexistence regime \cite{Georges92, Jarrell_Hubbard_1992, Rozenberg_Mott_1992, Zhang_Mott_1993, Rozenberg99, Bulla99,  Bluemer_PhD}.

\begin{figure}[tb]
\centering
\includegraphics{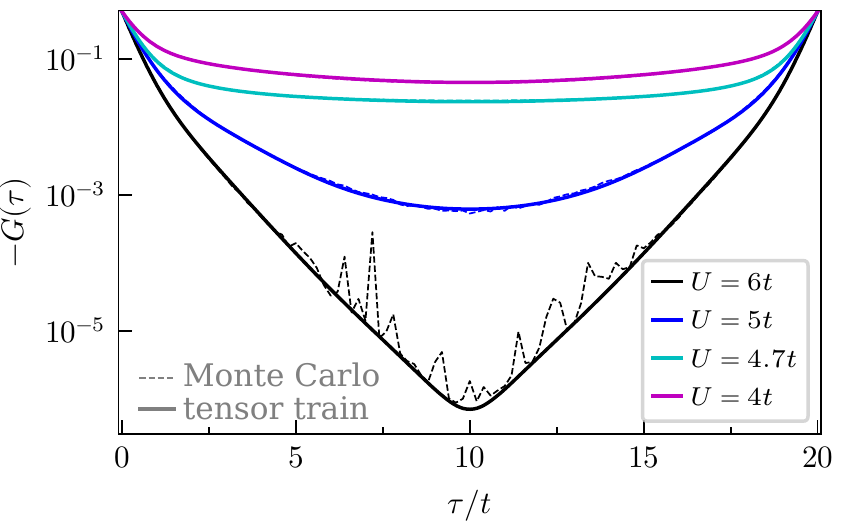}
\caption{
DMFT GF for representative interaction strengths $U$ at inverse temperature $\beta=20/t$. The dashed lines represent results obtained by hybridization expansion CT-QMC \cite{Werner06,Hafermann13,Gaenko17} (measured in imaginary time as described in \cite{Werner06}).
The full lines show results calculated within the tensor train approach.
The tensor rank necessary for obtaining accurate results depends on interaction strength and ranges from $90$ in the insulating regime to $200$ in the metallic regime.
Both results agree within their respective errors; differences between the two methods are visible at large $U$ due to the logarithmic scale.
}
\label{fig:DMFT_GF}
\end{figure}

We first study results for the GF obtained at different electron-electron interaction strength $U$ at temperature $\beta=20/t$.
The GF for representative values of $U$ is depicted in Fig.~\ref{fig:DMFT_GF}, as calculated by both standard CT-QMC and the tensor train approach.
Both methods agree within their respective errors. For $U \geq 5t$, the spectral weight $A(\omega=0) \sim \beta G(\beta/2)$ becomes strongly suppressed, indicating the opening of the Mott gap and the qualitative difference between metallic and insulating solutions. 
We emphasize that both methods sample the same diagrammatic perturbation expansion \cite{Werner06}, either by performing a stochastic random walk in diagram space or by calculating a tensor train approximation to the integrand at different orders.

Fig.~\ref{fig:DMFT_GF} demonstrates that the tensor train approach can provide results that are compatible with findings obtained within CT-QMC schemes, and therefore establishes tensor train based schemes as an alternative to Monte Carlo based impurity solvers.
As is evident from the data, the tensor train method does not suffer from stochastic noise. 
While it can generally be much more precise as compared to CT-QMC methods at similar computational cost, there are numerical aspects that influence the precision of the tensor train result. 
In particular, when decomposing the integrands for high hybridization orders where the parameter space is vast, we found that the quality of the tensor train approximation can become sensitive to the details of the pivots that are chosen. This is especially the cases when the integrand is essentially zero, or when only a small part of the parameter space contributes to the integral (see Sec.~\ref{sec:hyb+TT} for more details). 
These cases require a careful analysis of the results obtained by the tensor train method, which in practice limits the feasibility of our current implementation of higher precision results for high hybridization orders. The situation is not unlike the one with ergodicity issues in Monte Carlo, where the choice of an initial state or of a few early moves may prevent the simulation from exploring the entirety of phase space.
Further investigations -- in particular with respect to how pivots are picked, how the inversion of the pivot matrix is performed, and what variables are used to represent the integrand at a specific hybridization order -- are left for future work.

\begin{figure*}[tb]
  \centering
  \includegraphics{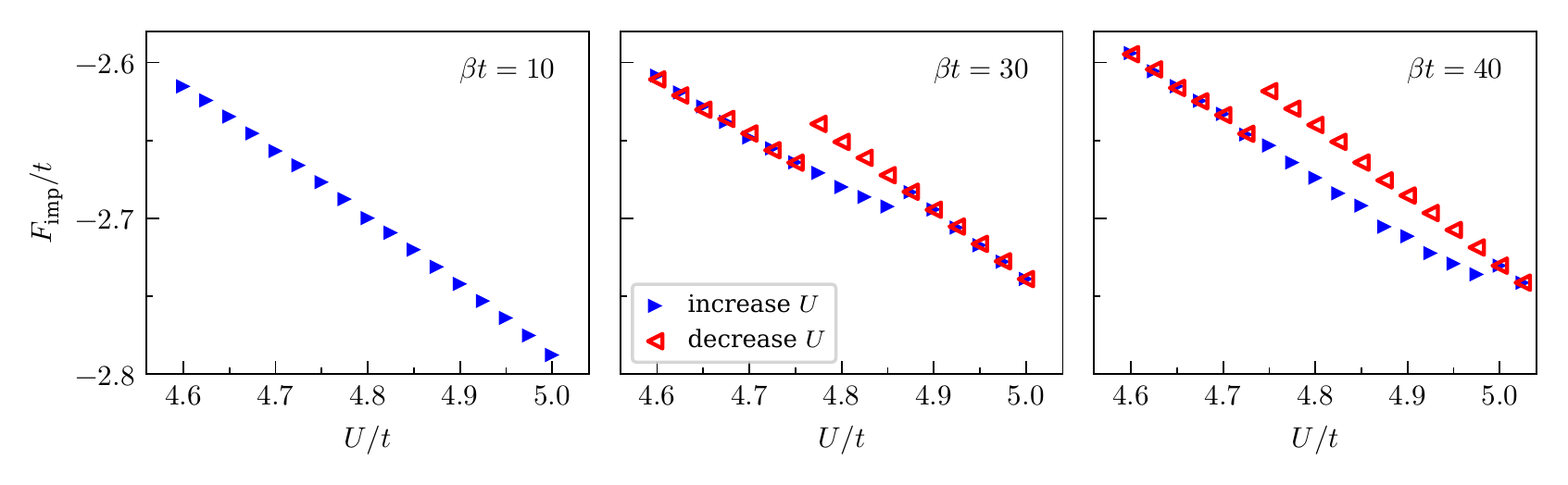}
  \caption{
      Impurity free energy $F_{\text{imp}}=-\log(Z)/\beta$, which includes the contribution from both spins, as a function of interaction strength $U$. The three panels correspond to three representative temperatures. 
      The metallic phase is found for small $U$, while large values of $U$ correspond to the insulating phase. 
      The two branches represent the DMFT results starting from the system in the metallic phase and successively increasing the interaction strength (indicated by the markers ${\color{blue}\blacktriangleright}$), 
      and DMFT results starting from the system in the insulating phase and successively increasing the interaction strength (indicated by the markers ${\color{red}\triangleleft}$), respectively. 
      For intermediate values of $U\sim4.8t$, the two branches separate, indicating the existence of two co-existing phases. 
      Upon comparing results for different fixed ranks, we estimate our relative error to be below $2\cdot10^{-3}$.
  }
  \label{fig:Phase_transition}
\end{figure*}
We now focus on the thermodynamic properties of the system at the metal-to-insulator transition.
The first-order phase transition and the coexistence regime between metallic and insulating solutions in the single-site DMFT has been investigated in great detail \cite{Bluemer_PhD,Bulla99,Schlipf99,Rozenberg99,Krauth00,Bulla01,Joo01}.

Thermodynamic quantities are directly accessible from the tensor train formalism, since the partition function can be obtained using Eq.~(\ref{eq:integral_expression_Z}). This is in contrast to CT-QMC, where diagrams are sampled with the probability that they contribute to the partition function, but an overall partition function normalization factor is typically not accessible (see Ref.~\cite{Troyer03} and Ref.~\cite{Gull11_RMP} Sec. X.E on quantum Wang--Landau algorithms for sampling this normalization in `bare' expansions and Ref.~\cite{Cohen15} for normalizing to the hypervolume of the time integral in renormalized/inchworm perturbation theory), and in contrast to the Hirsch--Fye algorithm \cite{Hirsch86}, where thermodynamic integration was used to delineate the phase boundary \cite{Bluemer_PhD}.

Fig.~\ref{fig:Phase_transition} shows the impurity free energy $F_{\text{imp}}=-\log(Z)/\beta$, as a function of interaction strength $U$ for three representative temperatures, calculated directly from the partition function Eq.~(\ref{eq:integral_expression_Z}) using the tensor train methodology. 
The impurity free energy is closely related to the lattice free energy $F_{\text{lattice}} = F_{\text{imp}} + \sum_\sigma\int_0^\beta \Delta^2(\tau)/t^2 d\tau$  \cite{Georges_Strongly_1992, Kotliar_Landau_2000}.
We perform the underlying DMFT calculation starting from two different reference systems: 
(i) the metallic system at small $U$ where we successively increase the interaction strength and 
(ii) the insulating system at large $U$ where we successively decrease the interaction strength.
Below the critical temperature of $\beta \sim 15/t$, we find two coexisting solutions with differing free energy.
This implies a coexistence regime where both metallic and insulating solutions can be stabilized, the extent of which increases with decreasing temperature. 

\section{Conclusion}\label{sec:conclusion}
In conclusion, we presented a method for solving strongly correlated equilibrium quantum impurity problems by expressing the terms in a diagrammatic series expansion by an approximate tensor train form, so that integration over internal degrees of freedom becomes tractable. We tested the method on a typical problem in the field: the single-site Anderson impurity model, as it appears in the context of dynamical mean-field theory.
Since the method is based on the hybridization expansion underlying commonly used CT-QMC algorithms, much of the knowledge and experience is directly transferable.
We showed that CT-QMC and tensor train methods lead to consistent results.
However, the tensor train results were more precise than CT-QMC results for the problems studied here, and do not suffer from any meaningful level of stochastic noise.
We were able to converge the tensor train approximation in all cases shown here to a level of accuracy that is very costly to achieve in CT-QMC.
Moreover, we showed that in contrast to CT-QMC approaches, the tensor train methodology allows for direct access to the partition function and thermodynamic properties.

Tensor train methods show great promise as solvers for equilibrium quantum impurity problems, since the limitations of the tensor train approximation are expected to be very different from that of CT-QMC. 
Nevertheless, since we employed a hybridization expansion \cite{Werner06_Kondo}, the overall computational cost scales exponentially with the number of impurity orbitals.
While we find that the current methodology does have limitations (such as, in certain cases, the selection of near-singular pivots for the tensor cross interpolation that may lead to imprecise tensor train approximations), we believe that further research into the numerics of tensor train approximations will overcome these issues.

Future application that promise to substantially increase the parameter space of impurity problems that can be solved reliably include partial summation techniques using the `inchworm' \cite{Cohen15,Eidelstein20,Cai20,kim_pseudoparticle_2022} or `bold' methodologies \cite{Prokofev07,Prokofev08,Gull10_Bold,cohen_numerically_2013,cohen_greens_2014,cohen_greens_2014-1}, multi-orbital impurity systems with frustrations, complex interactions, and general off-diagonal hybridizations \cite{Werner06_Kondo,Haule07,Eidelstein20,Li22}, and steady-state real-time \cite{Erpenbeck_Quantum_2022} and non-equilibrium \cite{Muhlbacher_Real_2008,Schiro_Real_2009,Werner_Diagrammatic_2009,Cohen15} formulations. 
Other potential applications include high-precision analytic continuation schemes \cite{Fei21}, interaction expansion series \cite{Rubtsov05,Gull08} including equilibrium and bold-line nonequilibrium methods, as well as other types of `diagrammatic' and `continuous-time' formalisms that are traditionally treated evaluated using Monte Carlo methods \cite{vanHoucke12,Gull11_RMP}.

\section*{Acknowledgements}
    A.E. was funded by the Deutsche Forschungsgemeinschaft (DFG, German Research Foundation) -- 45364484.
    E.G. and W.-T. L. were supported by the Department of Energy via DE-SC-0022088.
    T.B. was funded by the Department of Energy via DE-SC0020347.
    This material is based upon work supported by the U.S. Department of Energy, Office of Science, Office of Advanced Scientific Computing Research and Office of Basic Energy Sciences, Scientific Discovery through Advanced Computing (SciDAC) program under Award Number DE-SC0022088. 
    This research used resources of the National Energy Research Scientific Computing Center, a DOE Office of Science User Facility supported by the Office of Science of the U.S. Department of Energy under Contract No. DE-AC02-05CH11231 using NERSC award BES-ERCAP0021805.
    G.C. acknowledges support by the Israel Science Foundation (Grants No. 2902/21 and 218/19) and by the PAZY foundation (Grant No. 318/78).
    X.W. acknowledges support from the Plan France 2030 ANR-22-PETQ-0007 "EPIQ".
    The Flatiron Institute is a division of the Simons Foundation.

\section*{Author Contributions}
A.E. and W.-T. L. contributed equally to this paper.

\bibliography{refs}

\begin{thebibliography}{87}%
\makeatletter
\providecommand \@ifxundefined [1]{%
 \@ifx{#1\undefined}
}%
\providecommand \@ifnum [1]{%
 \ifnum #1\expandafter \@firstoftwo
 \else \expandafter \@secondoftwo
 \fi
}%
\providecommand \@ifx [1]{%
 \ifx #1\expandafter \@firstoftwo
 \else \expandafter \@secondoftwo
 \fi
}%
\providecommand \natexlab [1]{#1}%
\providecommand \enquote  [1]{``#1''}%
\providecommand \bibnamefont  [1]{#1}%
\providecommand \bibfnamefont [1]{#1}%
\providecommand \citenamefont [1]{#1}%
\providecommand \href@noop [0]{\@secondoftwo}%
\providecommand \href [0]{\begingroup \@sanitize@url \@href}%
\providecommand \@href[1]{\@@startlink{#1}\@@href}%
\providecommand \@@href[1]{\endgroup#1\@@endlink}%
\providecommand \@sanitize@url [0]{\catcode `\\12\catcode `\$12\catcode
  `\&12\catcode `\#12\catcode `\^12\catcode `\_12\catcode `\%12\relax}%
\providecommand \@@startlink[1]{}%
\providecommand \@@endlink[0]{}%
\providecommand \url  [0]{\begingroup\@sanitize@url \@url }%
\providecommand \@url [1]{\endgroup\@href {#1}{\urlprefix }}%
\providecommand \urlprefix  [0]{URL }%
\providecommand \Eprint [0]{\href }%
\providecommand \doibase [0]{https://doi.org/}%
\providecommand \selectlanguage [0]{\@gobble}%
\providecommand \bibinfo  [0]{\@secondoftwo}%
\providecommand \bibfield  [0]{\@secondoftwo}%
\providecommand \translation [1]{[#1]}%
\providecommand \BibitemOpen [0]{}%
\providecommand \bibitemStop [0]{}%
\providecommand \bibitemNoStop [0]{.\EOS\space}%
\providecommand \EOS [0]{\spacefactor3000\relax}%
\providecommand \BibitemShut  [1]{\csname bibitem#1\endcsname}%
\let\auto@bib@innerbib\@empty
\bibitem [{\citenamefont {Anderson}(1961)}]{Anderson61}%
  \BibitemOpen
  \bibfield  {author} {\bibinfo {author} {\bibfnamefont {P.~W.}\ \bibnamefont
  {Anderson}},\ }\href {https://doi.org/10.1103/PhysRev.124.41} {\bibfield
  {journal} {\bibinfo  {journal} {Phys. Rev.}\ }\textbf {\bibinfo {volume}
  {124}},\ \bibinfo {pages} {41} (\bibinfo {year} {1961})}\BibitemShut
  {NoStop}%
\bibitem [{\citenamefont {Hanson}\ \emph {et~al.}(2007)\citenamefont {Hanson},
  \citenamefont {Kouwenhoven}, \citenamefont {Petta}, \citenamefont {Tarucha},\
  and\ \citenamefont {Vandersypen}}]{Hanson07}%
  \BibitemOpen
  \bibfield  {author} {\bibinfo {author} {\bibfnamefont {R.}~\bibnamefont
  {Hanson}}, \bibinfo {author} {\bibfnamefont {L.~P.}\ \bibnamefont
  {Kouwenhoven}}, \bibinfo {author} {\bibfnamefont {J.~R.}\ \bibnamefont
  {Petta}}, \bibinfo {author} {\bibfnamefont {S.}~\bibnamefont {Tarucha}},\
  and\ \bibinfo {author} {\bibfnamefont {L.~M.~K.}\ \bibnamefont
  {Vandersypen}},\ }\href {https://doi.org/10.1103/RevModPhys.79.1217}
  {\bibfield  {journal} {\bibinfo  {journal} {Rev. Mod. Phys.}\ }\textbf
  {\bibinfo {volume} {79}},\ \bibinfo {pages} {1217} (\bibinfo {year}
  {2007})}\BibitemShut {NoStop}%
\bibitem [{\citenamefont {Brako}\ and\ \citenamefont {Newns}(1981)}]{Brako81}%
  \BibitemOpen
  \bibfield  {author} {\bibinfo {author} {\bibfnamefont {R.}~\bibnamefont
  {Brako}}\ and\ \bibinfo {author} {\bibfnamefont {D.~M.}\ \bibnamefont
  {Newns}},\ }\href {https://doi.org/10.1088/0022-3719/14/21/023} {\bibfield
  {journal} {\bibinfo  {journal} {Journal of Physics C: Solid State Physics}\
  }\textbf {\bibinfo {volume} {14}},\ \bibinfo {pages} {3065} (\bibinfo {year}
  {1981})}\BibitemShut {NoStop}%
\bibitem [{\citenamefont {Langreth}\ and\ \citenamefont
  {Nordlander}(1991)}]{Langreth91}%
  \BibitemOpen
  \bibfield  {author} {\bibinfo {author} {\bibfnamefont {D.~C.}\ \bibnamefont
  {Langreth}}\ and\ \bibinfo {author} {\bibfnamefont {P.}~\bibnamefont
  {Nordlander}},\ }\href {https://doi.org/10.1103/PhysRevB.43.2541} {\bibfield
  {journal} {\bibinfo  {journal} {Phys. Rev. B}\ }\textbf {\bibinfo {volume}
  {43}},\ \bibinfo {pages} {2541} (\bibinfo {year} {1991})}\BibitemShut
  {NoStop}%
\bibitem [{\citenamefont {Georges}\ and\ \citenamefont
  {Kotliar}(1992)}]{Georges92}%
  \BibitemOpen
  \bibfield  {author} {\bibinfo {author} {\bibfnamefont {A.}~\bibnamefont
  {Georges}}\ and\ \bibinfo {author} {\bibfnamefont {G.}~\bibnamefont
  {Kotliar}},\ }\href {https://doi.org/10.1103/PhysRevB.45.6479} {\bibfield
  {journal} {\bibinfo  {journal} {Phys. Rev. B}\ }\textbf {\bibinfo {volume}
  {45}},\ \bibinfo {pages} {6479} (\bibinfo {year} {1992})}\BibitemShut
  {NoStop}%
\bibitem [{\citenamefont {Georges}\ \emph {et~al.}(1996)\citenamefont
  {Georges}, \citenamefont {Kotliar}, \citenamefont {Krauth},\ and\
  \citenamefont {Rozenberg}}]{Georges96}%
  \BibitemOpen
  \bibfield  {author} {\bibinfo {author} {\bibfnamefont {A.}~\bibnamefont
  {Georges}}, \bibinfo {author} {\bibfnamefont {G.}~\bibnamefont {Kotliar}},
  \bibinfo {author} {\bibfnamefont {W.}~\bibnamefont {Krauth}},\ and\ \bibinfo
  {author} {\bibfnamefont {M.~J.}\ \bibnamefont {Rozenberg}},\ }\href
  {https://doi.org/10.1103/RevModPhys.68.13} {\bibfield  {journal} {\bibinfo
  {journal} {Rev. Mod. Phys.}\ }\textbf {\bibinfo {volume} {68}},\ \bibinfo
  {pages} {13} (\bibinfo {year} {1996})}\BibitemShut {NoStop}%
\bibitem [{\citenamefont {Kotliar}\ \emph {et~al.}(2006)\citenamefont
  {Kotliar}, \citenamefont {Savrasov}, \citenamefont {Haule}, \citenamefont
  {Oudovenko}, \citenamefont {Parcollet},\ and\ \citenamefont
  {Marianetti}}]{Kotliar06}%
  \BibitemOpen
  \bibfield  {author} {\bibinfo {author} {\bibfnamefont {G.}~\bibnamefont
  {Kotliar}}, \bibinfo {author} {\bibfnamefont {S.~Y.}\ \bibnamefont
  {Savrasov}}, \bibinfo {author} {\bibfnamefont {K.}~\bibnamefont {Haule}},
  \bibinfo {author} {\bibfnamefont {V.~S.}\ \bibnamefont {Oudovenko}}, \bibinfo
  {author} {\bibfnamefont {O.}~\bibnamefont {Parcollet}},\ and\ \bibinfo
  {author} {\bibfnamefont {C.~A.}\ \bibnamefont {Marianetti}},\ }\href
  {https://doi.org/10.1103/RevModPhys.78.865} {\bibfield  {journal} {\bibinfo
  {journal} {Rev. Mod. Phys.}\ }\textbf {\bibinfo {volume} {78}},\ \bibinfo
  {pages} {865} (\bibinfo {year} {2006})}\BibitemShut {NoStop}%
\bibitem [{\citenamefont {Zgid}\ and\ \citenamefont {Gull}(2017)}]{Zgid17}%
  \BibitemOpen
  \bibfield  {author} {\bibinfo {author} {\bibfnamefont {D.}~\bibnamefont
  {Zgid}}\ and\ \bibinfo {author} {\bibfnamefont {E.}~\bibnamefont {Gull}},\
  }\href {https://doi.org/10.1088/1367-2630/aa5d34} {\bibfield  {journal}
  {\bibinfo  {journal} {New Journal of Physics}\ }\textbf {\bibinfo {volume}
  {19}},\ \bibinfo {pages} {023047} (\bibinfo {year} {2017})}\BibitemShut
  {NoStop}%
\bibitem [{\citenamefont {Rubtsov}\ \emph {et~al.}(2005)\citenamefont
  {Rubtsov}, \citenamefont {Savkin},\ and\ \citenamefont
  {Lichtenstein}}]{Rubtsov05}%
  \BibitemOpen
  \bibfield  {author} {\bibinfo {author} {\bibfnamefont {A.~N.}\ \bibnamefont
  {Rubtsov}}, \bibinfo {author} {\bibfnamefont {V.~V.}\ \bibnamefont
  {Savkin}},\ and\ \bibinfo {author} {\bibfnamefont {A.~I.}\ \bibnamefont
  {Lichtenstein}},\ }\href {https://doi.org/10.1103/PhysRevB.72.035122}
  {\bibfield  {journal} {\bibinfo  {journal} {Phys. Rev. B}\ }\textbf {\bibinfo
  {volume} {72}},\ \bibinfo {pages} {035122} (\bibinfo {year}
  {2005})}\BibitemShut {NoStop}%
\bibitem [{\citenamefont {Werner}\ \emph {et~al.}(2006)\citenamefont {Werner},
  \citenamefont {Comanac}, \citenamefont {de' Medici}, \citenamefont {Troyer},\
  and\ \citenamefont {Millis}}]{Werner06}%
  \BibitemOpen
  \bibfield  {author} {\bibinfo {author} {\bibfnamefont {P.}~\bibnamefont
  {Werner}}, \bibinfo {author} {\bibfnamefont {A.}~\bibnamefont {Comanac}},
  \bibinfo {author} {\bibfnamefont {L.}~\bibnamefont {de' Medici}}, \bibinfo
  {author} {\bibfnamefont {M.}~\bibnamefont {Troyer}},\ and\ \bibinfo {author}
  {\bibfnamefont {A.~J.}\ \bibnamefont {Millis}},\ }\href
  {https://doi.org/10.1103/PhysRevLett.97.076405} {\bibfield  {journal}
  {\bibinfo  {journal} {Phys. Rev. Lett.}\ }\textbf {\bibinfo {volume} {97}},\
  \bibinfo {pages} {076405} (\bibinfo {year} {2006})}\BibitemShut {NoStop}%
\bibitem [{\citenamefont {Gull}\ \emph {et~al.}(2008)\citenamefont {Gull},
  \citenamefont {Werner}, \citenamefont {Parcollet},\ and\ \citenamefont
  {Troyer}}]{Gull08}%
  \BibitemOpen
  \bibfield  {author} {\bibinfo {author} {\bibfnamefont {E.}~\bibnamefont
  {Gull}}, \bibinfo {author} {\bibfnamefont {P.}~\bibnamefont {Werner}},
  \bibinfo {author} {\bibfnamefont {O.}~\bibnamefont {Parcollet}},\ and\
  \bibinfo {author} {\bibfnamefont {M.}~\bibnamefont {Troyer}},\ }\href
  {https://doi.org/10.1209/0295-5075/82/57003} {\bibfield  {journal} {\bibinfo
  {journal} {Europhysics Letters}\ }\textbf {\bibinfo {volume} {82}},\ \bibinfo
  {pages} {57003} (\bibinfo {year} {2008})}\BibitemShut {NoStop}%
\bibitem [{\citenamefont {Gull}\ \emph {et~al.}(2011)\citenamefont {Gull},
  \citenamefont {Millis}, \citenamefont {Lichtenstein}, \citenamefont
  {Rubtsov}, \citenamefont {Troyer},\ and\ \citenamefont
  {Werner}}]{Gull11_RMP}%
  \BibitemOpen
  \bibfield  {author} {\bibinfo {author} {\bibfnamefont {E.}~\bibnamefont
  {Gull}}, \bibinfo {author} {\bibfnamefont {A.~J.}\ \bibnamefont {Millis}},
  \bibinfo {author} {\bibfnamefont {A.~I.}\ \bibnamefont {Lichtenstein}},
  \bibinfo {author} {\bibfnamefont {A.~N.}\ \bibnamefont {Rubtsov}}, \bibinfo
  {author} {\bibfnamefont {M.}~\bibnamefont {Troyer}},\ and\ \bibinfo {author}
  {\bibfnamefont {P.}~\bibnamefont {Werner}},\ }\href
  {https://doi.org/10.1103/RevModPhys.83.349} {\bibfield  {journal} {\bibinfo
  {journal} {Rev. Mod. Phys.}\ }\textbf {\bibinfo {volume} {83}},\ \bibinfo
  {pages} {349} (\bibinfo {year} {2011})}\BibitemShut {NoStop}%
\bibitem [{\citenamefont {Maier}\ \emph {et~al.}(2005)\citenamefont {Maier},
  \citenamefont {Jarrell}, \citenamefont {Pruschke},\ and\ \citenamefont
  {Hettler}}]{Maier05}%
  \BibitemOpen
  \bibfield  {author} {\bibinfo {author} {\bibfnamefont {T.}~\bibnamefont
  {Maier}}, \bibinfo {author} {\bibfnamefont {M.}~\bibnamefont {Jarrell}},
  \bibinfo {author} {\bibfnamefont {T.}~\bibnamefont {Pruschke}},\ and\
  \bibinfo {author} {\bibfnamefont {M.~H.}\ \bibnamefont {Hettler}},\ }\href
  {https://doi.org/10.1103/RevModPhys.77.1027} {\bibfield  {journal} {\bibinfo
  {journal} {Rev. Mod. Phys.}\ }\textbf {\bibinfo {volume} {77}},\ \bibinfo
  {pages} {1027} (\bibinfo {year} {2005})}\BibitemShut {NoStop}%
\bibitem [{\citenamefont {Werner}\ and\ \citenamefont
  {Millis}(2006)}]{Werner06_Kondo}%
  \BibitemOpen
  \bibfield  {author} {\bibinfo {author} {\bibfnamefont {P.}~\bibnamefont
  {Werner}}\ and\ \bibinfo {author} {\bibfnamefont {A.~J.}\ \bibnamefont
  {Millis}},\ }\href {https://doi.org/10.1103/PhysRevB.74.155107} {\bibfield
  {journal} {\bibinfo  {journal} {Phys. Rev. B}\ }\textbf {\bibinfo {volume}
  {74}},\ \bibinfo {pages} {155107} (\bibinfo {year} {2006})}\BibitemShut
  {NoStop}%
\bibitem [{\citenamefont {Haule}(2007)}]{Haule07}%
  \BibitemOpen
  \bibfield  {author} {\bibinfo {author} {\bibfnamefont {K.}~\bibnamefont
  {Haule}},\ }\href {https://doi.org/10.1103/PhysRevB.75.155113} {\bibfield
  {journal} {\bibinfo  {journal} {Phys. Rev. B}\ }\textbf {\bibinfo {volume}
  {75}},\ \bibinfo {pages} {155113} (\bibinfo {year} {2007})}\BibitemShut
  {NoStop}%
\bibitem [{\citenamefont {M\"uhlbacher}\ and\ \citenamefont
  {Rabani}(2008{\natexlab{a}})}]{Muhlbacher08}%
  \BibitemOpen
  \bibfield  {author} {\bibinfo {author} {\bibfnamefont {L.}~\bibnamefont
  {M\"uhlbacher}}\ and\ \bibinfo {author} {\bibfnamefont {E.}~\bibnamefont
  {Rabani}},\ }\href {https://doi.org/10.1103/PhysRevLett.100.176403}
  {\bibfield  {journal} {\bibinfo  {journal} {Phys. Rev. Lett.}\ }\textbf
  {\bibinfo {volume} {100}},\ \bibinfo {pages} {176403} (\bibinfo {year}
  {2008}{\natexlab{a}})}\BibitemShut {NoStop}%
\bibitem [{\citenamefont {Gull}\ \emph
  {et~al.}(2010{\natexlab{a}})\citenamefont {Gull}, \citenamefont {Reichman},\
  and\ \citenamefont {Millis}}]{Gull10}%
  \BibitemOpen
  \bibfield  {author} {\bibinfo {author} {\bibfnamefont {E.}~\bibnamefont
  {Gull}}, \bibinfo {author} {\bibfnamefont {D.~R.}\ \bibnamefont {Reichman}},\
  and\ \bibinfo {author} {\bibfnamefont {A.~J.}\ \bibnamefont {Millis}},\
  }\href {https://doi.org/10.1103/PhysRevB.82.075109} {\bibfield  {journal}
  {\bibinfo  {journal} {Phys. Rev. B}\ }\textbf {\bibinfo {volume} {82}},\
  \bibinfo {pages} {075109} (\bibinfo {year} {2010}{\natexlab{a}})}\BibitemShut
  {NoStop}%
\bibitem [{\citenamefont {Cohen}\ \emph {et~al.}(2015)\citenamefont {Cohen},
  \citenamefont {Gull}, \citenamefont {Reichman},\ and\ \citenamefont
  {Millis}}]{Cohen15}%
  \BibitemOpen
  \bibfield  {author} {\bibinfo {author} {\bibfnamefont {G.}~\bibnamefont
  {Cohen}}, \bibinfo {author} {\bibfnamefont {E.}~\bibnamefont {Gull}},
  \bibinfo {author} {\bibfnamefont {D.~R.}\ \bibnamefont {Reichman}},\ and\
  \bibinfo {author} {\bibfnamefont {A.~J.}\ \bibnamefont {Millis}},\ }\href
  {https://doi.org/10.1103/PhysRevLett.115.266802} {\bibfield  {journal}
  {\bibinfo  {journal} {Phys. Rev. Lett.}\ }\textbf {\bibinfo {volume} {115}},\
  \bibinfo {pages} {266802} (\bibinfo {year} {2015})}\BibitemShut {NoStop}%
\bibitem [{\citenamefont {Gunacker}\ \emph {et~al.}(2015)\citenamefont
  {Gunacker}, \citenamefont {Wallerberger}, \citenamefont {Gull}, \citenamefont
  {Hausoel}, \citenamefont {Sangiovanni},\ and\ \citenamefont
  {Held}}]{Gunacker15}%
  \BibitemOpen
  \bibfield  {author} {\bibinfo {author} {\bibfnamefont {P.}~\bibnamefont
  {Gunacker}}, \bibinfo {author} {\bibfnamefont {M.}~\bibnamefont
  {Wallerberger}}, \bibinfo {author} {\bibfnamefont {E.}~\bibnamefont {Gull}},
  \bibinfo {author} {\bibfnamefont {A.}~\bibnamefont {Hausoel}}, \bibinfo
  {author} {\bibfnamefont {G.}~\bibnamefont {Sangiovanni}},\ and\ \bibinfo
  {author} {\bibfnamefont {K.}~\bibnamefont {Held}},\ }\href
  {https://doi.org/10.1103/PhysRevB.92.155102} {\bibfield  {journal} {\bibinfo
  {journal} {Phys. Rev. B}\ }\textbf {\bibinfo {volume} {92}},\ \bibinfo
  {pages} {155102} (\bibinfo {year} {2015})}\BibitemShut {NoStop}%
\bibitem [{\citenamefont {Eidelstein}\ \emph {et~al.}(2020)\citenamefont
  {Eidelstein}, \citenamefont {Gull},\ and\ \citenamefont
  {Cohen}}]{Eidelstein20}%
  \BibitemOpen
  \bibfield  {author} {\bibinfo {author} {\bibfnamefont {E.}~\bibnamefont
  {Eidelstein}}, \bibinfo {author} {\bibfnamefont {E.}~\bibnamefont {Gull}},\
  and\ \bibinfo {author} {\bibfnamefont {G.}~\bibnamefont {Cohen}},\ }\href
  {https://doi.org/10.1103/PhysRevLett.124.206405} {\bibfield  {journal}
  {\bibinfo  {journal} {Phys. Rev. Lett.}\ }\textbf {\bibinfo {volume} {124}},\
  \bibinfo {pages} {206405} (\bibinfo {year} {2020})}\BibitemShut {NoStop}%
\bibitem [{\citenamefont {Li}\ \emph {et~al.}(2022)\citenamefont {Li},
  \citenamefont {Yu}, \citenamefont {Gull},\ and\ \citenamefont
  {Cohen}}]{Li22}%
  \BibitemOpen
  \bibfield  {author} {\bibinfo {author} {\bibfnamefont {J.}~\bibnamefont
  {Li}}, \bibinfo {author} {\bibfnamefont {Y.}~\bibnamefont {Yu}}, \bibinfo
  {author} {\bibfnamefont {E.}~\bibnamefont {Gull}},\ and\ \bibinfo {author}
  {\bibfnamefont {G.}~\bibnamefont {Cohen}},\ }\href
  {https://doi.org/10.1103/PhysRevB.105.165133} {\bibfield  {journal} {\bibinfo
   {journal} {Phys. Rev. B}\ }\textbf {\bibinfo {volume} {105}},\ \bibinfo
  {pages} {165133} (\bibinfo {year} {2022})}\BibitemShut {NoStop}%
\bibitem [{\citenamefont {Bauer}\ \emph {et~al.}(2011)\citenamefont {Bauer},
  \citenamefont {Carr}, \citenamefont {Evertz}, \citenamefont {Feiguin},
  \citenamefont {Freire}, \citenamefont {Fuchs}, \citenamefont {Gamper},
  \citenamefont {Gukelberger}, \citenamefont {Gull}, \citenamefont {Guertler},
  \citenamefont {Hehn}, \citenamefont {Igarashi}, \citenamefont {Isakov},
  \citenamefont {Koop}, \citenamefont {Ma}, \citenamefont {Mates},
  \citenamefont {Matsuo}, \citenamefont {Parcollet}, \citenamefont
  {Pawłowski}, \citenamefont {Picon}, \citenamefont {Pollet}, \citenamefont
  {Santos}, \citenamefont {Scarola}, \citenamefont {Schollwöck}, \citenamefont
  {Silva}, \citenamefont {Surer}, \citenamefont {Todo}, \citenamefont {Trebst},
  \citenamefont {Troyer}, \citenamefont {Wall}, \citenamefont {Werner},\ and\
  \citenamefont {Wessel}}]{Bauer11}%
  \BibitemOpen
  \bibfield  {author} {\bibinfo {author} {\bibfnamefont {B.}~\bibnamefont
  {Bauer}}, \bibinfo {author} {\bibfnamefont {L.~D.}\ \bibnamefont {Carr}},
  \bibinfo {author} {\bibfnamefont {H.~G.}\ \bibnamefont {Evertz}}, \bibinfo
  {author} {\bibfnamefont {A.}~\bibnamefont {Feiguin}}, \bibinfo {author}
  {\bibfnamefont {J.}~\bibnamefont {Freire}}, \bibinfo {author} {\bibfnamefont
  {S.}~\bibnamefont {Fuchs}}, \bibinfo {author} {\bibfnamefont
  {L.}~\bibnamefont {Gamper}}, \bibinfo {author} {\bibfnamefont
  {J.}~\bibnamefont {Gukelberger}}, \bibinfo {author} {\bibfnamefont
  {E.}~\bibnamefont {Gull}}, \bibinfo {author} {\bibfnamefont {S.}~\bibnamefont
  {Guertler}}, \bibinfo {author} {\bibfnamefont {A.}~\bibnamefont {Hehn}},
  \bibinfo {author} {\bibfnamefont {R.}~\bibnamefont {Igarashi}}, \bibinfo
  {author} {\bibfnamefont {S.~V.}\ \bibnamefont {Isakov}}, \bibinfo {author}
  {\bibfnamefont {D.}~\bibnamefont {Koop}}, \bibinfo {author} {\bibfnamefont
  {P.~N.}\ \bibnamefont {Ma}}, \bibinfo {author} {\bibfnamefont
  {P.}~\bibnamefont {Mates}}, \bibinfo {author} {\bibfnamefont
  {H.}~\bibnamefont {Matsuo}}, \bibinfo {author} {\bibfnamefont
  {O.}~\bibnamefont {Parcollet}}, \bibinfo {author} {\bibfnamefont
  {G.}~\bibnamefont {Pawłowski}}, \bibinfo {author} {\bibfnamefont {J.~D.}\
  \bibnamefont {Picon}}, \bibinfo {author} {\bibfnamefont {L.}~\bibnamefont
  {Pollet}}, \bibinfo {author} {\bibfnamefont {E.}~\bibnamefont {Santos}},
  \bibinfo {author} {\bibfnamefont {V.~W.}\ \bibnamefont {Scarola}}, \bibinfo
  {author} {\bibfnamefont {U.}~\bibnamefont {Schollwöck}}, \bibinfo {author}
  {\bibfnamefont {C.}~\bibnamefont {Silva}}, \bibinfo {author} {\bibfnamefont
  {B.}~\bibnamefont {Surer}}, \bibinfo {author} {\bibfnamefont
  {S.}~\bibnamefont {Todo}}, \bibinfo {author} {\bibfnamefont {S.}~\bibnamefont
  {Trebst}}, \bibinfo {author} {\bibfnamefont {M.}~\bibnamefont {Troyer}},
  \bibinfo {author} {\bibfnamefont {M.~L.}\ \bibnamefont {Wall}}, \bibinfo
  {author} {\bibfnamefont {P.}~\bibnamefont {Werner}},\ and\ \bibinfo {author}
  {\bibfnamefont {S.}~\bibnamefont {Wessel}},\ }\href
  {https://doi.org/10.1088/1742-5468/2011/05/P05001} {\bibfield  {journal}
  {\bibinfo  {journal} {Journal of Statistical Mechanics: Theory and
  Experiment}\ }\textbf {\bibinfo {volume} {2011}},\ \bibinfo {pages} {P05001}
  (\bibinfo {year} {2011})}\BibitemShut {NoStop}%
\bibitem [{\citenamefont {Shinaoka}\ \emph {et~al.}(2014)\citenamefont
  {Shinaoka}, \citenamefont {Dolfi}, \citenamefont {Troyer},\ and\
  \citenamefont {Werner}}]{Shinaoka14}%
  \BibitemOpen
  \bibfield  {author} {\bibinfo {author} {\bibfnamefont {H.}~\bibnamefont
  {Shinaoka}}, \bibinfo {author} {\bibfnamefont {M.}~\bibnamefont {Dolfi}},
  \bibinfo {author} {\bibfnamefont {M.}~\bibnamefont {Troyer}},\ and\ \bibinfo
  {author} {\bibfnamefont {P.}~\bibnamefont {Werner}},\ }\href
  {https://doi.org/10.1088/1742-5468/2014/06/P06012} {\bibfield  {journal}
  {\bibinfo  {journal} {Journal of Statistical Mechanics: Theory and
  Experiment}\ }\textbf {\bibinfo {volume} {2014}},\ \bibinfo {pages} {P06012}
  (\bibinfo {year} {2014})}\BibitemShut {NoStop}%
\bibitem [{\citenamefont {Parcollet}\ \emph {et~al.}(2015)\citenamefont
  {Parcollet}, \citenamefont {Ferrero}, \citenamefont {Ayral}, \citenamefont
  {Hafermann}, \citenamefont {Krivenko}, \citenamefont {Messio},\ and\
  \citenamefont {Seth}}]{TRIQS15}%
  \BibitemOpen
  \bibfield  {author} {\bibinfo {author} {\bibfnamefont {O.}~\bibnamefont
  {Parcollet}}, \bibinfo {author} {\bibfnamefont {M.}~\bibnamefont {Ferrero}},
  \bibinfo {author} {\bibfnamefont {T.}~\bibnamefont {Ayral}}, \bibinfo
  {author} {\bibfnamefont {H.}~\bibnamefont {Hafermann}}, \bibinfo {author}
  {\bibfnamefont {I.}~\bibnamefont {Krivenko}}, \bibinfo {author}
  {\bibfnamefont {L.}~\bibnamefont {Messio}},\ and\ \bibinfo {author}
  {\bibfnamefont {P.}~\bibnamefont {Seth}},\ }\href
  {https://doi.org/https://doi.org/10.1016/j.cpc.2015.04.023} {\bibfield
  {journal} {\bibinfo  {journal} {Computer Physics Communications}\ }\textbf
  {\bibinfo {volume} {196}},\ \bibinfo {pages} {398} (\bibinfo {year}
  {2015})}\BibitemShut {NoStop}%
\bibitem [{\citenamefont {Seth}\ \emph {et~al.}(2016)\citenamefont {Seth},
  \citenamefont {Krivenko}, \citenamefont {Ferrero},\ and\ \citenamefont
  {Parcollet}}]{Seth16}%
  \BibitemOpen
  \bibfield  {author} {\bibinfo {author} {\bibfnamefont {P.}~\bibnamefont
  {Seth}}, \bibinfo {author} {\bibfnamefont {I.}~\bibnamefont {Krivenko}},
  \bibinfo {author} {\bibfnamefont {M.}~\bibnamefont {Ferrero}},\ and\ \bibinfo
  {author} {\bibfnamefont {O.}~\bibnamefont {Parcollet}},\ }\href
  {https://doi.org/https://doi.org/10.1016/j.cpc.2015.10.023} {\bibfield
  {journal} {\bibinfo  {journal} {Computer Physics Communications}\ }\textbf
  {\bibinfo {volume} {200}},\ \bibinfo {pages} {274} (\bibinfo {year}
  {2016})}\BibitemShut {NoStop}%
\bibitem [{\citenamefont {Gaenko}\ \emph {et~al.}(2017)\citenamefont {Gaenko},
  \citenamefont {Antipov}, \citenamefont {Carcassi}, \citenamefont {Chen},
  \citenamefont {Chen}, \citenamefont {Dong}, \citenamefont {Gamper},
  \citenamefont {Gukelberger}, \citenamefont {Igarashi}, \citenamefont
  {Iskakov}, \citenamefont {Könz}, \citenamefont {LeBlanc}, \citenamefont
  {Levy}, \citenamefont {Ma}, \citenamefont {Paki}, \citenamefont {Shinaoka},
  \citenamefont {Todo}, \citenamefont {Troyer},\ and\ \citenamefont
  {Gull}}]{Gaenko17}%
  \BibitemOpen
  \bibfield  {author} {\bibinfo {author} {\bibfnamefont {A.}~\bibnamefont
  {Gaenko}}, \bibinfo {author} {\bibfnamefont {A.}~\bibnamefont {Antipov}},
  \bibinfo {author} {\bibfnamefont {G.}~\bibnamefont {Carcassi}}, \bibinfo
  {author} {\bibfnamefont {T.}~\bibnamefont {Chen}}, \bibinfo {author}
  {\bibfnamefont {X.}~\bibnamefont {Chen}}, \bibinfo {author} {\bibfnamefont
  {Q.}~\bibnamefont {Dong}}, \bibinfo {author} {\bibfnamefont {L.}~\bibnamefont
  {Gamper}}, \bibinfo {author} {\bibfnamefont {J.}~\bibnamefont {Gukelberger}},
  \bibinfo {author} {\bibfnamefont {R.}~\bibnamefont {Igarashi}}, \bibinfo
  {author} {\bibfnamefont {S.}~\bibnamefont {Iskakov}}, \bibinfo {author}
  {\bibfnamefont {M.}~\bibnamefont {Könz}}, \bibinfo {author} {\bibfnamefont
  {J.}~\bibnamefont {LeBlanc}}, \bibinfo {author} {\bibfnamefont
  {R.}~\bibnamefont {Levy}}, \bibinfo {author} {\bibfnamefont {P.}~\bibnamefont
  {Ma}}, \bibinfo {author} {\bibfnamefont {J.}~\bibnamefont {Paki}}, \bibinfo
  {author} {\bibfnamefont {H.}~\bibnamefont {Shinaoka}}, \bibinfo {author}
  {\bibfnamefont {S.}~\bibnamefont {Todo}}, \bibinfo {author} {\bibfnamefont
  {M.}~\bibnamefont {Troyer}},\ and\ \bibinfo {author} {\bibfnamefont
  {E.}~\bibnamefont {Gull}},\ }\href
  {https://doi.org/https://doi.org/10.1016/j.cpc.2016.12.009} {\bibfield
  {journal} {\bibinfo  {journal} {Computer Physics Communications}\ }\textbf
  {\bibinfo {volume} {213}},\ \bibinfo {pages} {235} (\bibinfo {year}
  {2017})}\BibitemShut {NoStop}%
\bibitem [{\citenamefont {Shinaoka}\ \emph {et~al.}(2017)\citenamefont
  {Shinaoka}, \citenamefont {Gull},\ and\ \citenamefont {Werner}}]{Shinaoka17}%
  \BibitemOpen
  \bibfield  {author} {\bibinfo {author} {\bibfnamefont {H.}~\bibnamefont
  {Shinaoka}}, \bibinfo {author} {\bibfnamefont {E.}~\bibnamefont {Gull}},\
  and\ \bibinfo {author} {\bibfnamefont {P.}~\bibnamefont {Werner}},\ }\href
  {https://doi.org/https://doi.org/10.1016/j.cpc.2017.01.003} {\bibfield
  {journal} {\bibinfo  {journal} {Computer Physics Communications}\ }\textbf
  {\bibinfo {volume} {215}},\ \bibinfo {pages} {128} (\bibinfo {year}
  {2017})}\BibitemShut {NoStop}%
\bibitem [{\citenamefont {Yue}\ \emph {et~al.}(2019)\citenamefont {Yue},
  \citenamefont {Wang}, \citenamefont {Otsuki},\ and\ \citenamefont
  {Dai}}]{Yue19}%
  \BibitemOpen
  \bibfield  {author} {\bibinfo {author} {\bibfnamefont {C.}~\bibnamefont
  {Yue}}, \bibinfo {author} {\bibfnamefont {Y.}~\bibnamefont {Wang}}, \bibinfo
  {author} {\bibfnamefont {J.}~\bibnamefont {Otsuki}},\ and\ \bibinfo {author}
  {\bibfnamefont {X.}~\bibnamefont {Dai}},\ }\href
  {https://doi.org/https://doi.org/10.1016/j.cpc.2018.10.025} {\bibfield
  {journal} {\bibinfo  {journal} {Computer Physics Communications}\ }\textbf
  {\bibinfo {volume} {236}},\ \bibinfo {pages} {135} (\bibinfo {year}
  {2019})}\BibitemShut {NoStop}%
\bibitem [{\citenamefont {Wallerberger}\ \emph {et~al.}(2019)\citenamefont
  {Wallerberger}, \citenamefont {Hausoel}, \citenamefont {Gunacker},
  \citenamefont {Kowalski}, \citenamefont {Parragh}, \citenamefont {Goth},
  \citenamefont {Held},\ and\ \citenamefont {Sangiovanni}}]{Wallerberger19}%
  \BibitemOpen
  \bibfield  {author} {\bibinfo {author} {\bibfnamefont {M.}~\bibnamefont
  {Wallerberger}}, \bibinfo {author} {\bibfnamefont {A.}~\bibnamefont
  {Hausoel}}, \bibinfo {author} {\bibfnamefont {P.}~\bibnamefont {Gunacker}},
  \bibinfo {author} {\bibfnamefont {A.}~\bibnamefont {Kowalski}}, \bibinfo
  {author} {\bibfnamefont {N.}~\bibnamefont {Parragh}}, \bibinfo {author}
  {\bibfnamefont {F.}~\bibnamefont {Goth}}, \bibinfo {author} {\bibfnamefont
  {K.}~\bibnamefont {Held}},\ and\ \bibinfo {author} {\bibfnamefont
  {G.}~\bibnamefont {Sangiovanni}},\ }\href
  {https://doi.org/https://doi.org/10.1016/j.cpc.2018.09.007} {\bibfield
  {journal} {\bibinfo  {journal} {Computer Physics Communications}\ }\textbf
  {\bibinfo {volume} {235}},\ \bibinfo {pages} {388} (\bibinfo {year}
  {2019})}\BibitemShut {NoStop}%
\bibitem [{\citenamefont {Shinaoka}\ \emph {et~al.}(2020)\citenamefont
  {Shinaoka}, \citenamefont {Nomura},\ and\ \citenamefont {Gull}}]{Shinaoka20}%
  \BibitemOpen
  \bibfield  {author} {\bibinfo {author} {\bibfnamefont {H.}~\bibnamefont
  {Shinaoka}}, \bibinfo {author} {\bibfnamefont {Y.}~\bibnamefont {Nomura}},\
  and\ \bibinfo {author} {\bibfnamefont {E.}~\bibnamefont {Gull}},\ }\href
  {https://doi.org/https://doi.org/10.1016/j.cpc.2019.06.016} {\bibfield
  {journal} {\bibinfo  {journal} {Computer Physics Communications}\ }\textbf
  {\bibinfo {volume} {252}},\ \bibinfo {pages} {106826} (\bibinfo {year}
  {2020})}\BibitemShut {NoStop}%
\bibitem [{\citenamefont {Caffarel}\ and\ \citenamefont
  {Krauth}(1994)}]{Caffarel_Exact_1994}%
  \BibitemOpen
  \bibfield  {author} {\bibinfo {author} {\bibfnamefont {M.}~\bibnamefont
  {Caffarel}}\ and\ \bibinfo {author} {\bibfnamefont {W.}~\bibnamefont
  {Krauth}},\ }\href {https://doi.org/10.1103/PhysRevLett.72.1545} {\bibfield
  {journal} {\bibinfo  {journal} {Phys. Rev. Lett.}\ }\textbf {\bibinfo
  {volume} {72}},\ \bibinfo {pages} {1545} (\bibinfo {year}
  {1994})}\BibitemShut {NoStop}%
\bibitem [{\citenamefont {Koch}\ \emph {et~al.}(2008)\citenamefont {Koch},
  \citenamefont {Sangiovanni},\ and\ \citenamefont {Gunnarsson}}]{Koch08}%
  \BibitemOpen
  \bibfield  {author} {\bibinfo {author} {\bibfnamefont {E.}~\bibnamefont
  {Koch}}, \bibinfo {author} {\bibfnamefont {G.}~\bibnamefont {Sangiovanni}},\
  and\ \bibinfo {author} {\bibfnamefont {O.}~\bibnamefont {Gunnarsson}},\
  }\href {https://doi.org/10.1103/PhysRevB.78.115102} {\bibfield  {journal}
  {\bibinfo  {journal} {Phys. Rev. B}\ }\textbf {\bibinfo {volume} {78}},\
  \bibinfo {pages} {115102} (\bibinfo {year} {2008})}\BibitemShut {NoStop}%
\bibitem [{\citenamefont {Mejuto-Zaera}\ \emph {et~al.}(2020)\citenamefont
  {Mejuto-Zaera}, \citenamefont {Zepeda-N\'u\~nez}, \citenamefont {Lindsey},
  \citenamefont {Tubman}, \citenamefont {Whaley},\ and\ \citenamefont
  {Lin}}]{MejutoZaera20}%
  \BibitemOpen
  \bibfield  {author} {\bibinfo {author} {\bibfnamefont {C.}~\bibnamefont
  {Mejuto-Zaera}}, \bibinfo {author} {\bibfnamefont {L.}~\bibnamefont
  {Zepeda-N\'u\~nez}}, \bibinfo {author} {\bibfnamefont {M.}~\bibnamefont
  {Lindsey}}, \bibinfo {author} {\bibfnamefont {N.}~\bibnamefont {Tubman}},
  \bibinfo {author} {\bibfnamefont {B.}~\bibnamefont {Whaley}},\ and\ \bibinfo
  {author} {\bibfnamefont {L.}~\bibnamefont {Lin}},\ }\href
  {https://doi.org/10.1103/PhysRevB.101.035143} {\bibfield  {journal} {\bibinfo
   {journal} {Phys. Rev. B}\ }\textbf {\bibinfo {volume} {101}},\ \bibinfo
  {pages} {035143} (\bibinfo {year} {2020})}\BibitemShut {NoStop}%
\bibitem [{\citenamefont {Hirsch}\ and\ \citenamefont {Fye}(1986)}]{Hirsch86}%
  \BibitemOpen
  \bibfield  {author} {\bibinfo {author} {\bibfnamefont {J.~E.}\ \bibnamefont
  {Hirsch}}\ and\ \bibinfo {author} {\bibfnamefont {R.~M.}\ \bibnamefont
  {Fye}},\ }\href {https://doi.org/10.1103/PhysRevLett.56.2521} {\bibfield
  {journal} {\bibinfo  {journal} {Phys. Rev. Lett.}\ }\textbf {\bibinfo
  {volume} {56}},\ \bibinfo {pages} {2521} (\bibinfo {year}
  {1986})}\BibitemShut {NoStop}%
\bibitem [{\citenamefont {Blankenbecler}\ \emph {et~al.}(1981)\citenamefont
  {Blankenbecler}, \citenamefont {Scalapino},\ and\ \citenamefont
  {Sugar}}]{Blankenbecler81}%
  \BibitemOpen
  \bibfield  {author} {\bibinfo {author} {\bibfnamefont {R.}~\bibnamefont
  {Blankenbecler}}, \bibinfo {author} {\bibfnamefont {D.~J.}\ \bibnamefont
  {Scalapino}},\ and\ \bibinfo {author} {\bibfnamefont {R.~L.}\ \bibnamefont
  {Sugar}},\ }\href {https://doi.org/10.1103/PhysRevD.24.2278} {\bibfield
  {journal} {\bibinfo  {journal} {Phys. Rev. D}\ }\textbf {\bibinfo {volume}
  {24}},\ \bibinfo {pages} {2278} (\bibinfo {year} {1981})}\BibitemShut
  {NoStop}%
\bibitem [{\citenamefont {Troyer}\ \emph {et~al.}(2003)\citenamefont {Troyer},
  \citenamefont {Wessel},\ and\ \citenamefont {Alet}}]{Troyer03}%
  \BibitemOpen
  \bibfield  {author} {\bibinfo {author} {\bibfnamefont {M.}~\bibnamefont
  {Troyer}}, \bibinfo {author} {\bibfnamefont {S.}~\bibnamefont {Wessel}},\
  and\ \bibinfo {author} {\bibfnamefont {F.}~\bibnamefont {Alet}},\ }\href
  {https://doi.org/10.1103/PhysRevLett.90.120201} {\bibfield  {journal}
  {\bibinfo  {journal} {Phys. Rev. Lett.}\ }\textbf {\bibinfo {volume} {90}},\
  \bibinfo {pages} {120201} (\bibinfo {year} {2003})}\BibitemShut {NoStop}%
\bibitem [{\citenamefont {Prokof'ev}\ \emph {et~al.}(1996)\citenamefont
  {Prokof'ev}, \citenamefont {Svistunov},\ and\ \citenamefont
  {Tupitsyn}}]{Prokofev96}%
  \BibitemOpen
  \bibfield  {author} {\bibinfo {author} {\bibfnamefont {N.~V.}\ \bibnamefont
  {Prokof'ev}}, \bibinfo {author} {\bibfnamefont {B.~V.}\ \bibnamefont
  {Svistunov}},\ and\ \bibinfo {author} {\bibfnamefont {I.~S.}\ \bibnamefont
  {Tupitsyn}},\ }\href {https://doi.org/10.1134/1.567243} {\bibfield  {journal}
  {\bibinfo  {journal} {JETP Letters}\ }\textbf {\bibinfo {volume} {64}},\
  \bibinfo {pages} {911} (\bibinfo {year} {1996})}\BibitemShut {NoStop}%
\bibitem [{\citenamefont {Prokof'ev}\ \emph {et~al.}(1998)\citenamefont
  {Prokof'ev}, \citenamefont {Svistunov},\ and\ \citenamefont
  {Tupitsyn}}]{Prokofev98A}%
  \BibitemOpen
  \bibfield  {author} {\bibinfo {author} {\bibfnamefont {N.~V.}\ \bibnamefont
  {Prokof'ev}}, \bibinfo {author} {\bibfnamefont {B.~V.}\ \bibnamefont
  {Svistunov}},\ and\ \bibinfo {author} {\bibfnamefont {I.~S.}\ \bibnamefont
  {Tupitsyn}},\ }\href {http://arxiv.org/abs/cond-mat/9703200} {\bibfield
  {journal} {\bibinfo  {journal} {JETP Sov. Phys.}\ }\textbf {\bibinfo {volume}
  {87}},\ \bibinfo {pages} {310} (\bibinfo {year} {1998})}\BibitemShut
  {NoStop}%
\bibitem [{\citenamefont {Prokof'ev}\ and\ \citenamefont
  {Svistunov}(1998)}]{Prokofev98B}%
  \BibitemOpen
  \bibfield  {author} {\bibinfo {author} {\bibfnamefont {N.~V.}\ \bibnamefont
  {Prokof'ev}}\ and\ \bibinfo {author} {\bibfnamefont {B.~V.}\ \bibnamefont
  {Svistunov}},\ }\href {https://doi.org/10.1103/PhysRevLett.81.2514}
  {\bibfield  {journal} {\bibinfo  {journal} {Phys. Rev. Lett.}\ }\textbf
  {\bibinfo {volume} {81}},\ \bibinfo {pages} {2514} (\bibinfo {year}
  {1998})}\BibitemShut {NoStop}%
\bibitem [{\citenamefont {Ma\ifmmode~\check{c}\else \v{c}\fi{}ek}\ \emph
  {et~al.}(2020)\citenamefont {Ma\ifmmode~\check{c}\else \v{c}\fi{}ek},
  \citenamefont {Dumitrescu}, \citenamefont {Bertrand}, \citenamefont {Triggs},
  \citenamefont {Parcollet},\ and\ \citenamefont {Waintal}}]{Macek20}%
  \BibitemOpen
  \bibfield  {author} {\bibinfo {author} {\bibfnamefont {M.}~\bibnamefont
  {Ma\ifmmode~\check{c}\else \v{c}\fi{}ek}}, \bibinfo {author} {\bibfnamefont
  {P.~T.}\ \bibnamefont {Dumitrescu}}, \bibinfo {author} {\bibfnamefont
  {C.}~\bibnamefont {Bertrand}}, \bibinfo {author} {\bibfnamefont
  {B.}~\bibnamefont {Triggs}}, \bibinfo {author} {\bibfnamefont
  {O.}~\bibnamefont {Parcollet}},\ and\ \bibinfo {author} {\bibfnamefont
  {X.}~\bibnamefont {Waintal}},\ }\href
  {https://doi.org/10.1103/PhysRevLett.125.047702} {\bibfield  {journal}
  {\bibinfo  {journal} {Phys. Rev. Lett.}\ }\textbf {\bibinfo {volume} {125}},\
  \bibinfo {pages} {047702} (\bibinfo {year} {2020})}\BibitemShut {NoStop}%
\bibitem [{\citenamefont {Bertrand}\ \emph {et~al.}(2021)\citenamefont
  {Bertrand}, \citenamefont {Bauernfeind}, \citenamefont {Dumitrescu},
  \citenamefont {Ma\ifmmode~\check{c}\else \v{c}\fi{}ek}, \citenamefont
  {Waintal},\ and\ \citenamefont {Parcollet}}]{Corentin21}%
  \BibitemOpen
  \bibfield  {author} {\bibinfo {author} {\bibfnamefont {C.}~\bibnamefont
  {Bertrand}}, \bibinfo {author} {\bibfnamefont {D.}~\bibnamefont
  {Bauernfeind}}, \bibinfo {author} {\bibfnamefont {P.~T.}\ \bibnamefont
  {Dumitrescu}}, \bibinfo {author} {\bibfnamefont {M.}~\bibnamefont
  {Ma\ifmmode~\check{c}\else \v{c}\fi{}ek}}, \bibinfo {author} {\bibfnamefont
  {X.}~\bibnamefont {Waintal}},\ and\ \bibinfo {author} {\bibfnamefont
  {O.}~\bibnamefont {Parcollet}},\ }\href
  {https://doi.org/10.1103/PhysRevB.103.155104} {\bibfield  {journal} {\bibinfo
   {journal} {Phys. Rev. B}\ }\textbf {\bibinfo {volume} {103}},\ \bibinfo
  {pages} {155104} (\bibinfo {year} {2021})}\BibitemShut {NoStop}%
\bibitem [{\citenamefont {N{\'u}{\~n}ez-Fern{\'a}ndez}\ \emph
  {et~al.}(2022)\citenamefont {N{\'u}{\~n}ez-Fern{\'a}ndez}, \citenamefont
  {Jeannin}, \citenamefont {Dumitrescu}, \citenamefont {Kloss}, \citenamefont
  {Kaye}, \citenamefont {Parcollet},\ and\ \citenamefont
  {Waintal}}]{NunezFernandez22}%
  \BibitemOpen
  \bibfield  {author} {\bibinfo {author} {\bibfnamefont {Y.}~\bibnamefont
  {N{\'u}{\~n}ez-Fern{\'a}ndez}}, \bibinfo {author} {\bibfnamefont
  {M.}~\bibnamefont {Jeannin}}, \bibinfo {author} {\bibfnamefont {P.~T.}\
  \bibnamefont {Dumitrescu}}, \bibinfo {author} {\bibfnamefont
  {T.}~\bibnamefont {Kloss}}, \bibinfo {author} {\bibfnamefont
  {J.}~\bibnamefont {Kaye}}, \bibinfo {author} {\bibfnamefont {O.}~\bibnamefont
  {Parcollet}},\ and\ \bibinfo {author} {\bibfnamefont {X.}~\bibnamefont
  {Waintal}},\ }\href {https://doi.org/10.1103/PhysRevX.12.041018} {\bibfield
  {journal} {\bibinfo  {journal} {Physical Review X}\ }\textbf {\bibinfo
  {volume} {12}},\ \bibinfo {pages} {041018} (\bibinfo {year}
  {2022})}\BibitemShut {NoStop}%
\bibitem [{\citenamefont {Dolgov}\ and\ \citenamefont
  {Savostyanov}(2020)}]{dolgov_parallel_2020}%
  \BibitemOpen
  \bibfield  {author} {\bibinfo {author} {\bibfnamefont {S.}~\bibnamefont
  {Dolgov}}\ and\ \bibinfo {author} {\bibfnamefont {D.}~\bibnamefont
  {Savostyanov}},\ }\href {https://doi.org/10.1016/j.cpc.2019.106869}
  {\bibfield  {journal} {\bibinfo  {journal} {Computer Physics Communications}\
  }\textbf {\bibinfo {volume} {246}},\ \bibinfo {pages} {106869} (\bibinfo
  {year} {2020})}\BibitemShut {NoStop}%
\bibitem [{\citenamefont {Goreinov}(2008)}]{goreinov_cross_2008}%
  \BibitemOpen
  \bibfield  {author} {\bibinfo {author} {\bibfnamefont {S.~A.}\ \bibnamefont
  {Goreinov}},\ }\href {https://doi.org/10.1134/S106456240803023X} {\bibfield
  {journal} {\bibinfo  {journal} {Doklady Mathematics}\ }\textbf {\bibinfo
  {volume} {77}},\ \bibinfo {pages} {404} (\bibinfo {year} {2008})}\BibitemShut
  {NoStop}%
\bibitem [{\citenamefont {Savostyanov}(2014)}]{savostyanov2014}%
  \BibitemOpen
  \bibfield  {author} {\bibinfo {author} {\bibfnamefont {D.~V.}\ \bibnamefont
  {Savostyanov}},\ }\href {https://doi.org/10.1016/j.laa.2014.06.006}
  {\bibfield  {journal} {\bibinfo  {journal} {Linear Algebra and its
  Applications}\ }\textbf {\bibinfo {volume} {458}},\ \bibinfo {pages} {217}
  (\bibinfo {year} {2014})}\BibitemShut {NoStop}%
\bibitem [{\citenamefont {Oseledets}\ and\ \citenamefont
  {Tyrtyshnikov}(2010)}]{Oseledets2010}%
  \BibitemOpen
  \bibfield  {author} {\bibinfo {author} {\bibfnamefont {I.}~\bibnamefont
  {Oseledets}}\ and\ \bibinfo {author} {\bibfnamefont {E.}~\bibnamefont
  {Tyrtyshnikov}},\ }\href {https://doi.org/10.1016/j.laa.2009.07.024}
  {\bibfield  {journal} {\bibinfo  {journal} {Linear Algebra and its
  Applications}\ }\textbf {\bibinfo {volume} {432}},\ \bibinfo {pages} {70}
  (\bibinfo {year} {2010})}\BibitemShut {NoStop}%
\bibitem [{\citenamefont {White}(1992)}]{White92}%
  \BibitemOpen
  \bibfield  {author} {\bibinfo {author} {\bibfnamefont {S.~R.}\ \bibnamefont
  {White}},\ }\href {https://doi.org/10.1103/PhysRevLett.69.2863} {\bibfield
  {journal} {\bibinfo  {journal} {Phys. Rev. Lett.}\ }\textbf {\bibinfo
  {volume} {69}},\ \bibinfo {pages} {2863} (\bibinfo {year}
  {1992})}\BibitemShut {NoStop}%
\bibitem [{\citenamefont {Schollwöck}(2011)}]{Schollwoeck11}%
  \BibitemOpen
  \bibfield  {author} {\bibinfo {author} {\bibfnamefont {U.}~\bibnamefont
  {Schollwöck}},\ }\href
  {https://doi.org/https://doi.org/10.1016/j.aop.2010.09.012} {\bibfield
  {journal} {\bibinfo  {journal} {Annals of Physics}\ }\textbf {\bibinfo
  {volume} {326}},\ \bibinfo {pages} {96} (\bibinfo {year} {2011})},\ \bibinfo
  {note} {january 2011 Special Issue}\BibitemShut {NoStop}%
\bibitem [{\citenamefont {Cohen}\ and\ \citenamefont
  {Rabani}(2011)}]{cohen_memory_2011}%
  \BibitemOpen
  \bibfield  {author} {\bibinfo {author} {\bibfnamefont {G.}~\bibnamefont
  {Cohen}}\ and\ \bibinfo {author} {\bibfnamefont {E.}~\bibnamefont {Rabani}},\
  }\href {https://doi.org/10.1103/PhysRevB.84.075150} {\bibfield  {journal}
  {\bibinfo  {journal} {Physical Review B}\ }\textbf {\bibinfo {volume} {84}},\
  \bibinfo {pages} {075150} (\bibinfo {year} {2011})}\BibitemShut {NoStop}%
\bibitem [{\citenamefont {Keiter}\ and\ \citenamefont
  {Kimball}(1970)}]{Keiter70}%
  \BibitemOpen
  \bibfield  {author} {\bibinfo {author} {\bibfnamefont {H.}~\bibnamefont
  {Keiter}}\ and\ \bibinfo {author} {\bibfnamefont {J.~C.}\ \bibnamefont
  {Kimball}},\ }\href {https://doi.org/10.1103/PhysRevLett.25.672} {\bibfield
  {journal} {\bibinfo  {journal} {Phys. Rev. Lett.}\ }\textbf {\bibinfo
  {volume} {25}},\ \bibinfo {pages} {672} (\bibinfo {year} {1970})}\BibitemShut
  {NoStop}%
\bibitem [{\citenamefont {Pruschke}\ and\ \citenamefont
  {Grewe}(1989)}]{Pruschke89}%
  \BibitemOpen
  \bibfield  {author} {\bibinfo {author} {\bibfnamefont {T.}~\bibnamefont
  {Pruschke}}\ and\ \bibinfo {author} {\bibfnamefont {N.}~\bibnamefont
  {Grewe}},\ }\href {https://doi.org/10.1007/BF01311391} {\bibfield  {journal}
  {\bibinfo  {journal} {Z. Phys. B}\ }\textbf {\bibinfo {volume} {74}},\
  \bibinfo {pages} {439} (\bibinfo {year} {1989})}\BibitemShut {NoStop}%
\bibitem [{\citenamefont {M\"uhlbacher}\ and\ \citenamefont
  {Rabani}(2008{\natexlab{b}})}]{Muhlbacher_Real_2008}%
  \BibitemOpen
  \bibfield  {author} {\bibinfo {author} {\bibfnamefont {L.}~\bibnamefont
  {M\"uhlbacher}}\ and\ \bibinfo {author} {\bibfnamefont {E.}~\bibnamefont
  {Rabani}},\ }\href@noop {} {\bibfield  {journal} {\bibinfo  {journal} {Phys.
  Rev. Lett.}\ }\textbf {\bibinfo {volume} {100}},\ \bibinfo {pages} {176403}
  (\bibinfo {year} {2008}{\natexlab{b}})}\BibitemShut {NoStop}%
\bibitem [{\citenamefont {Gull}\ \emph
  {et~al.}(2010{\natexlab{b}})\citenamefont {Gull}, \citenamefont {Reichman},\
  and\ \citenamefont {Millis}}]{Gull10_Bold}%
  \BibitemOpen
  \bibfield  {author} {\bibinfo {author} {\bibfnamefont {E.}~\bibnamefont
  {Gull}}, \bibinfo {author} {\bibfnamefont {D.~R.}\ \bibnamefont {Reichman}},\
  and\ \bibinfo {author} {\bibfnamefont {A.~J.}\ \bibnamefont {Millis}},\
  }\href {https://doi.org/10.1103/PhysRevB.82.075109} {\bibfield  {journal}
  {\bibinfo  {journal} {Phys. Rev. B}\ }\textbf {\bibinfo {volume} {82}},\
  \bibinfo {pages} {075109} (\bibinfo {year} {2010}{\natexlab{b}})}\BibitemShut
  {NoStop}%
\bibitem [{\citenamefont {Cohen}\ \emph
  {et~al.}(2014{\natexlab{a}})\citenamefont {Cohen}, \citenamefont {Reichman},
  \citenamefont {Millis},\ and\ \citenamefont {Gull}}]{cohen_greens_2014}%
  \BibitemOpen
  \bibfield  {author} {\bibinfo {author} {\bibfnamefont {G.}~\bibnamefont
  {Cohen}}, \bibinfo {author} {\bibfnamefont {D.~R.}\ \bibnamefont {Reichman}},
  \bibinfo {author} {\bibfnamefont {A.~J.}\ \bibnamefont {Millis}},\ and\
  \bibinfo {author} {\bibfnamefont {E.}~\bibnamefont {Gull}},\ }\href
  {https://doi.org/10.1103/PhysRevB.89.115139} {\bibfield  {journal} {\bibinfo
  {journal} {Physical Review B}\ }\textbf {\bibinfo {volume} {89}},\ \bibinfo
  {pages} {115139} (\bibinfo {year} {2014}{\natexlab{a}})}\BibitemShut
  {NoStop}%
\bibitem [{\citenamefont {Cohen}\ \emph
  {et~al.}(2014{\natexlab{b}})\citenamefont {Cohen}, \citenamefont {Gull},
  \citenamefont {Reichman},\ and\ \citenamefont
  {Millis}}]{cohen_greens_2014-1}%
  \BibitemOpen
  \bibfield  {author} {\bibinfo {author} {\bibfnamefont {G.}~\bibnamefont
  {Cohen}}, \bibinfo {author} {\bibfnamefont {E.}~\bibnamefont {Gull}},
  \bibinfo {author} {\bibfnamefont {D.~R.}\ \bibnamefont {Reichman}},\ and\
  \bibinfo {author} {\bibfnamefont {A.~J.}\ \bibnamefont {Millis}},\ }\href
  {https://doi.org/10.1103/PhysRevLett.112.146802} {\bibfield  {journal}
  {\bibinfo  {journal} {Physical Review Letters}\ }\textbf {\bibinfo {volume}
  {112}},\ \bibinfo {pages} {146802} (\bibinfo {year}
  {2014}{\natexlab{b}})}\BibitemShut {NoStop}%
\bibitem [{\citenamefont {Prokof'ev}\ and\ \citenamefont
  {Svistunov}(2008)}]{Prokofev08}%
  \BibitemOpen
  \bibfield  {author} {\bibinfo {author} {\bibfnamefont {N.~V.}\ \bibnamefont
  {Prokof'ev}}\ and\ \bibinfo {author} {\bibfnamefont {B.~V.}\ \bibnamefont
  {Svistunov}},\ }\href {https://doi.org/10.1103/PhysRevB.77.125101} {\bibfield
   {journal} {\bibinfo  {journal} {Phys. Rev. B}\ }\textbf {\bibinfo {volume}
  {77}},\ \bibinfo {pages} {125101} (\bibinfo {year} {2008})}\BibitemShut
  {NoStop}%
\bibitem [{\citenamefont {Van~Houcke}\ \emph {et~al.}(2012)\citenamefont
  {Van~Houcke}, \citenamefont {Werner}, \citenamefont {Kozik}, \citenamefont
  {Prokof'ev}, \citenamefont {Svistunov}, \citenamefont {Ku}, \citenamefont
  {Sommer}, \citenamefont {Cheuk}, \citenamefont {Schirotzek},\ and\
  \citenamefont {Zwierlein}}]{vanHoucke12}%
  \BibitemOpen
  \bibfield  {author} {\bibinfo {author} {\bibfnamefont {K.}~\bibnamefont
  {Van~Houcke}}, \bibinfo {author} {\bibfnamefont {F.}~\bibnamefont {Werner}},
  \bibinfo {author} {\bibfnamefont {E.}~\bibnamefont {Kozik}}, \bibinfo
  {author} {\bibfnamefont {N.}~\bibnamefont {Prokof'ev}}, \bibinfo {author}
  {\bibfnamefont {B.}~\bibnamefont {Svistunov}}, \bibinfo {author}
  {\bibfnamefont {M.~J.~H.}\ \bibnamefont {Ku}}, \bibinfo {author}
  {\bibfnamefont {A.~T.}\ \bibnamefont {Sommer}}, \bibinfo {author}
  {\bibfnamefont {L.~W.}\ \bibnamefont {Cheuk}}, \bibinfo {author}
  {\bibfnamefont {A.}~\bibnamefont {Schirotzek}},\ and\ \bibinfo {author}
  {\bibfnamefont {M.~W.}\ \bibnamefont {Zwierlein}},\ }\href@noop {} {\bibfield
   {journal} {\bibinfo  {journal} {Nature Physics}\ }\textbf {\bibinfo {volume}
  {8}},\ \bibinfo {pages} {366} (\bibinfo {year} {2012})}\BibitemShut {NoStop}%
\bibitem [{\citenamefont {Oseledets}(2013)}]{oseledets2013}%
  \BibitemOpen
  \bibfield  {author} {\bibinfo {author} {\bibfnamefont {I.~V.}\ \bibnamefont
  {Oseledets}},\ }\href {https://doi.org/10.1007/s00365-012-9175-x} {\bibfield
  {journal} {\bibinfo  {journal} {Constructive Approximation}\ }\textbf
  {\bibinfo {volume} {37}},\ \bibinfo {pages} {1} (\bibinfo {year}
  {2013})}\BibitemShut {NoStop}%
\bibitem [{\citenamefont {Khoromskij}(2018)}]{khoromskij2018tensor}%
  \BibitemOpen
  \bibfield  {author} {\bibinfo {author} {\bibfnamefont {B.~N.}\ \bibnamefont
  {Khoromskij}},\ }\href@noop {} {\emph {\bibinfo {title} {Tensor Numerical
  Methods in Scientific Computing}}},\ Vol.~\bibinfo {volume} {19}\ (\bibinfo
  {publisher} {Walter de Gruyter GmbH \& Co KG},\ \bibinfo {year}
  {2018})\BibitemShut {NoStop}%
\bibitem [{\citenamefont {Tyrtyshnikov}(2000)}]{Tyrtyshnikov2000}%
  \BibitemOpen
  \bibfield  {author} {\bibinfo {author} {\bibfnamefont {E.}~\bibnamefont
  {Tyrtyshnikov}},\ }\href {https://doi.org/10.1007/s006070070031} {\bibfield
  {journal} {\bibinfo  {journal} {Computing}\ }\textbf {\bibinfo {volume}
  {64}},\ \bibinfo {pages} {367} (\bibinfo {year} {2000})}\BibitemShut
  {NoStop}%
\bibitem [{\citenamefont {Goreinov}\ \emph {et~al.}(1997)\citenamefont
  {Goreinov}, \citenamefont {Zamarashkin},\ and\ \citenamefont
  {Tyrtyshnikov}}]{goreinov_pseudo_1997}%
  \BibitemOpen
  \bibfield  {author} {\bibinfo {author} {\bibfnamefont {S.}~\bibnamefont
  {Goreinov}}, \bibinfo {author} {\bibfnamefont {N.}~\bibnamefont
  {Zamarashkin}},\ and\ \bibinfo {author} {\bibfnamefont {E.}~\bibnamefont
  {Tyrtyshnikov}},\ }\href@noop {} {\bibfield  {journal} {\bibinfo  {journal}
  {Mathematical Notes of the Academy of Sciences of the USSR}\ }\textbf
  {\bibinfo {volume} {62}},\ \bibinfo {pages} {515} (\bibinfo {year}
  {1997})}\BibitemShut {NoStop}%
\bibitem [{\citenamefont {Goreinov}\ \emph {et~al.}(2010)\citenamefont
  {Goreinov}, \citenamefont {Oseledets}, \citenamefont {Savostyanov},
  \citenamefont {Tyrtyshnikov},\ and\ \citenamefont
  {Zamarashkin}}]{Goreinov_Matrix_2010}%
  \BibitemOpen
  \bibfield  {author} {\bibinfo {author} {\bibfnamefont {S.~A.}\ \bibnamefont
  {Goreinov}}, \bibinfo {author} {\bibfnamefont {I.~V.}\ \bibnamefont
  {Oseledets}}, \bibinfo {author} {\bibfnamefont {D.~V.}\ \bibnamefont
  {Savostyanov}}, \bibinfo {author} {\bibfnamefont {E.~E.}\ \bibnamefont
  {Tyrtyshnikov}},\ and\ \bibinfo {author} {\bibfnamefont {N.~L.}\ \bibnamefont
  {Zamarashkin}},\ }in\ \href@noop {} {\emph {\bibinfo {booktitle} {Matrix
  Methods: Theory, Algorithms and Applications}}}\ (\bibinfo  {publisher}
  {World Scientific, Singapore},\ \bibinfo {year} {2010})\ p.\ \bibinfo {pages}
  {247–256}\BibitemShut {NoStop}%
\bibitem [{\citenamefont {Bruus}\ and\ \citenamefont
  {Flensberg}(2004)}]{Bruus_Many_2004}%
  \BibitemOpen
  \bibfield  {author} {\bibinfo {author} {\bibfnamefont {H.}~\bibnamefont
  {Bruus}}\ and\ \bibinfo {author} {\bibfnamefont {K.}~\bibnamefont
  {Flensberg}},\ }\href@noop {} {\emph {\bibinfo {title} {Many-Body Quantum
  Theory in Condensed Matter Physics: An Introduction}}}\ (\bibinfo
  {publisher} {Oxford Graduate Texts, Oxford},\ \bibinfo {year}
  {2004})\BibitemShut {NoStop}%
\bibitem [{\citenamefont {Haug}\ and\ \citenamefont
  {Jauho}(2008)}]{Haug_Qauntum_2008}%
  \BibitemOpen
  \bibfield  {author} {\bibinfo {author} {\bibfnamefont {H.~J.~W.}\
  \bibnamefont {Haug}}\ and\ \bibinfo {author} {\bibfnamefont {A.-P.}\
  \bibnamefont {Jauho}},\ }\href@noop {} {\emph {\bibinfo {title} {Quantum
  Kinetics in Transport and Optics of Semiconductors}}}\ (\bibinfo  {publisher}
  {Springer Series in Solid-State Sciences},\ \bibinfo {address} {Berlin,
  Heidelberg},\ \bibinfo {year} {2008})\BibitemShut {NoStop}%
\bibitem [{\citenamefont {Fei}\ \emph {et~al.}(2021)\citenamefont {Fei},
  \citenamefont {Yeh},\ and\ \citenamefont {Gull}}]{Fei21}%
  \BibitemOpen
  \bibfield  {author} {\bibinfo {author} {\bibfnamefont {J.}~\bibnamefont
  {Fei}}, \bibinfo {author} {\bibfnamefont {C.-N.}\ \bibnamefont {Yeh}},\ and\
  \bibinfo {author} {\bibfnamefont {E.}~\bibnamefont {Gull}},\ }\href
  {https://doi.org/10.1103/PhysRevLett.126.056402} {\bibfield  {journal}
  {\bibinfo  {journal} {Phys. Rev. Lett.}\ }\textbf {\bibinfo {volume} {126}},\
  \bibinfo {pages} {056402} (\bibinfo {year} {2021})}\BibitemShut {NoStop}%
\bibitem [{\citenamefont {Jarrell}(1992)}]{Jarrell_Hubbard_1992}%
  \BibitemOpen
  \bibfield  {author} {\bibinfo {author} {\bibfnamefont {M.}~\bibnamefont
  {Jarrell}},\ }\href {https://doi.org/10.1103/PhysRevLett.69.168} {\bibfield
  {journal} {\bibinfo  {journal} {Phys. Rev. Lett.}\ }\textbf {\bibinfo
  {volume} {69}},\ \bibinfo {pages} {168} (\bibinfo {year} {1992})}\BibitemShut
  {NoStop}%
\bibitem [{\citenamefont {Rozenberg}\ \emph {et~al.}(1992)\citenamefont
  {Rozenberg}, \citenamefont {Zhang},\ and\ \citenamefont
  {Kotliar}}]{Rozenberg_Mott_1992}%
  \BibitemOpen
  \bibfield  {author} {\bibinfo {author} {\bibfnamefont {M.~J.}\ \bibnamefont
  {Rozenberg}}, \bibinfo {author} {\bibfnamefont {X.~Y.}\ \bibnamefont
  {Zhang}},\ and\ \bibinfo {author} {\bibfnamefont {G.}~\bibnamefont
  {Kotliar}},\ }\href {https://doi.org/10.1103/PhysRevLett.69.1236} {\bibfield
  {journal} {\bibinfo  {journal} {Phys. Rev. Lett.}\ }\textbf {\bibinfo
  {volume} {69}},\ \bibinfo {pages} {1236} (\bibinfo {year}
  {1992})}\BibitemShut {NoStop}%
\bibitem [{\citenamefont {Zhang}\ \emph {et~al.}(1993)\citenamefont {Zhang},
  \citenamefont {Rozenberg},\ and\ \citenamefont {Kotliar}}]{Zhang_Mott_1993}%
  \BibitemOpen
  \bibfield  {author} {\bibinfo {author} {\bibfnamefont {X.~Y.}\ \bibnamefont
  {Zhang}}, \bibinfo {author} {\bibfnamefont {M.~J.}\ \bibnamefont
  {Rozenberg}},\ and\ \bibinfo {author} {\bibfnamefont {G.}~\bibnamefont
  {Kotliar}},\ }\href {https://doi.org/10.1103/PhysRevLett.70.1666} {\bibfield
  {journal} {\bibinfo  {journal} {Phys. Rev. Lett.}\ }\textbf {\bibinfo
  {volume} {70}},\ \bibinfo {pages} {1666} (\bibinfo {year}
  {1993})}\BibitemShut {NoStop}%
\bibitem [{\citenamefont {Rozenberg}\ \emph {et~al.}(1999)\citenamefont
  {Rozenberg}, \citenamefont {Chitra},\ and\ \citenamefont
  {Kotliar}}]{Rozenberg99}%
  \BibitemOpen
  \bibfield  {author} {\bibinfo {author} {\bibfnamefont {M.~J.}\ \bibnamefont
  {Rozenberg}}, \bibinfo {author} {\bibfnamefont {R.}~\bibnamefont {Chitra}},\
  and\ \bibinfo {author} {\bibfnamefont {G.}~\bibnamefont {Kotliar}},\ }\href
  {https://doi.org/10.1103/PhysRevLett.83.3498} {\bibfield  {journal} {\bibinfo
   {journal} {Phys. Rev. Lett.}\ }\textbf {\bibinfo {volume} {83}},\ \bibinfo
  {pages} {3498} (\bibinfo {year} {1999})}\BibitemShut {NoStop}%
\bibitem [{\citenamefont {Bulla}(1999)}]{Bulla99}%
  \BibitemOpen
  \bibfield  {author} {\bibinfo {author} {\bibfnamefont {R.}~\bibnamefont
  {Bulla}},\ }\href {https://doi.org/10.1103/PhysRevLett.83.136} {\bibfield
  {journal} {\bibinfo  {journal} {Phys. Rev. Lett.}\ }\textbf {\bibinfo
  {volume} {83}},\ \bibinfo {pages} {136} (\bibinfo {year} {1999})}\BibitemShut
  {NoStop}%
\bibitem [{\citenamefont {Bl\"umer}(2003)}]{Bluemer_PhD}%
  \BibitemOpen
  \bibfield  {author} {\bibinfo {author} {\bibfnamefont {N.}~\bibnamefont
  {Bl\"umer}},\ }\href@noop {} {\emph {\bibinfo {title} {Metal-Insulator
  Transition and Optical Conductivity in High Dimensions}}}\ (\bibinfo
  {publisher} {Shaker Verlag},\ \bibinfo {address} {Aachen},\ \bibinfo {year}
  {2003})\BibitemShut {NoStop}%
\bibitem [{\citenamefont {Metzner}\ and\ \citenamefont
  {Vollhardt}(1989)}]{Metzner_Correlated_1989}%
  \BibitemOpen
  \bibfield  {author} {\bibinfo {author} {\bibfnamefont {W.}~\bibnamefont
  {Metzner}}\ and\ \bibinfo {author} {\bibfnamefont {D.}~\bibnamefont
  {Vollhardt}},\ }\href {https://doi.org/10.1103/PhysRevLett.62.324} {\bibfield
   {journal} {\bibinfo  {journal} {Phys. Rev. Lett.}\ }\textbf {\bibinfo
  {volume} {62}},\ \bibinfo {pages} {324} (\bibinfo {year} {1989})}\BibitemShut
  {NoStop}%
\bibitem [{\citenamefont {Eckstein}\ \emph {et~al.}(2005)\citenamefont
  {Eckstein}, \citenamefont {Kollar}, \citenamefont {Byczuk},\ and\
  \citenamefont {Vollhardt}}]{Eckstein_Hopping_2005}%
  \BibitemOpen
  \bibfield  {author} {\bibinfo {author} {\bibfnamefont {M.}~\bibnamefont
  {Eckstein}}, \bibinfo {author} {\bibfnamefont {M.}~\bibnamefont {Kollar}},
  \bibinfo {author} {\bibfnamefont {K.}~\bibnamefont {Byczuk}},\ and\ \bibinfo
  {author} {\bibfnamefont {D.}~\bibnamefont {Vollhardt}},\ }\href
  {https://doi.org/10.1103/PhysRevB.71.235119} {\bibfield  {journal} {\bibinfo
  {journal} {Phys. Rev. B}\ }\textbf {\bibinfo {volume} {71}},\ \bibinfo
  {pages} {235119} (\bibinfo {year} {2005})}\BibitemShut {NoStop}%
\bibitem [{\citenamefont {Hafermann}\ \emph {et~al.}(2013)\citenamefont
  {Hafermann}, \citenamefont {Werner},\ and\ \citenamefont
  {Gull}}]{Hafermann13}%
  \BibitemOpen
  \bibfield  {author} {\bibinfo {author} {\bibfnamefont {H.}~\bibnamefont
  {Hafermann}}, \bibinfo {author} {\bibfnamefont {P.}~\bibnamefont {Werner}},\
  and\ \bibinfo {author} {\bibfnamefont {E.}~\bibnamefont {Gull}},\ }\href
  {https://doi.org/https://doi.org/10.1016/j.cpc.2012.12.013} {\bibfield
  {journal} {\bibinfo  {journal} {Computer Physics Communications}\ }\textbf
  {\bibinfo {volume} {184}},\ \bibinfo {pages} {1280} (\bibinfo {year}
  {2013})}\BibitemShut {NoStop}%
\bibitem [{\citenamefont {Schlipf}\ \emph {et~al.}(1999)\citenamefont
  {Schlipf}, \citenamefont {Jarrell}, \citenamefont {van Dongen}, \citenamefont
  {Bl\"umer}, \citenamefont {Kehrein}, \citenamefont {Pruschke},\ and\
  \citenamefont {Vollhardt}}]{Schlipf99}%
  \BibitemOpen
  \bibfield  {author} {\bibinfo {author} {\bibfnamefont {J.}~\bibnamefont
  {Schlipf}}, \bibinfo {author} {\bibfnamefont {M.}~\bibnamefont {Jarrell}},
  \bibinfo {author} {\bibfnamefont {P.~G.~J.}\ \bibnamefont {van Dongen}},
  \bibinfo {author} {\bibfnamefont {N.}~\bibnamefont {Bl\"umer}}, \bibinfo
  {author} {\bibfnamefont {S.}~\bibnamefont {Kehrein}}, \bibinfo {author}
  {\bibfnamefont {T.}~\bibnamefont {Pruschke}},\ and\ \bibinfo {author}
  {\bibfnamefont {D.}~\bibnamefont {Vollhardt}},\ }\href
  {https://doi.org/10.1103/PhysRevLett.82.4890} {\bibfield  {journal} {\bibinfo
   {journal} {Phys. Rev. Lett.}\ }\textbf {\bibinfo {volume} {82}},\ \bibinfo
  {pages} {4890} (\bibinfo {year} {1999})}\BibitemShut {NoStop}%
\bibitem [{\citenamefont {Krauth}(2000)}]{Krauth00}%
  \BibitemOpen
  \bibfield  {author} {\bibinfo {author} {\bibfnamefont {W.}~\bibnamefont
  {Krauth}},\ }\href {https://doi.org/10.1103/PhysRevB.62.6860} {\bibfield
  {journal} {\bibinfo  {journal} {Phys. Rev. B}\ }\textbf {\bibinfo {volume}
  {62}},\ \bibinfo {pages} {6860} (\bibinfo {year} {2000})}\BibitemShut
  {NoStop}%
\bibitem [{\citenamefont {Bulla}\ \emph {et~al.}(2001)\citenamefont {Bulla},
  \citenamefont {Costi},\ and\ \citenamefont {Vollhardt}}]{Bulla01}%
  \BibitemOpen
  \bibfield  {author} {\bibinfo {author} {\bibfnamefont {R.}~\bibnamefont
  {Bulla}}, \bibinfo {author} {\bibfnamefont {T.~A.}\ \bibnamefont {Costi}},\
  and\ \bibinfo {author} {\bibfnamefont {D.}~\bibnamefont {Vollhardt}},\ }\href
  {https://doi.org/10.1103/PhysRevB.64.045103} {\bibfield  {journal} {\bibinfo
  {journal} {Phys. Rev. B}\ }\textbf {\bibinfo {volume} {64}},\ \bibinfo
  {pages} {045103} (\bibinfo {year} {2001})}\BibitemShut {NoStop}%
\bibitem [{\citenamefont {Joo}\ and\ \citenamefont {Oudovenko}(2001)}]{Joo01}%
  \BibitemOpen
  \bibfield  {author} {\bibinfo {author} {\bibfnamefont {J.}~\bibnamefont
  {Joo}}\ and\ \bibinfo {author} {\bibfnamefont {V.}~\bibnamefont
  {Oudovenko}},\ }\href {https://doi.org/10.1103/PhysRevB.64.193102} {\bibfield
   {journal} {\bibinfo  {journal} {Phys. Rev. B}\ }\textbf {\bibinfo {volume}
  {64}},\ \bibinfo {pages} {193102} (\bibinfo {year} {2001})}\BibitemShut
  {NoStop}%
\bibitem [{\citenamefont {Georges}\ \emph {et~al.}(1992)\citenamefont
  {Georges}, \citenamefont {Kotliar},\ and\ \citenamefont
  {Si}}]{Georges_Strongly_1992}%
  \BibitemOpen
  \bibfield  {author} {\bibinfo {author} {\bibfnamefont {A.}~\bibnamefont
  {Georges}}, \bibinfo {author} {\bibfnamefont {G.}~\bibnamefont {Kotliar}},\
  and\ \bibinfo {author} {\bibfnamefont {Q.}~\bibnamefont {Si}},\ }\href
  {https://doi.org/10.1142/S0217979292000426} {\bibfield  {journal} {\bibinfo
  {journal} {Int. J. Mod. Phys B}\ }\textbf {\bibinfo {volume} {06}},\ \bibinfo
  {pages} {705} (\bibinfo {year} {1992})}\BibitemShut {NoStop}%
\bibitem [{\citenamefont {Kotliar}\ \emph {et~al.}(2000)\citenamefont
  {Kotliar}, \citenamefont {Lange},\ and\ \citenamefont
  {Rozenberg}}]{Kotliar_Landau_2000}%
  \BibitemOpen
  \bibfield  {author} {\bibinfo {author} {\bibfnamefont {G.}~\bibnamefont
  {Kotliar}}, \bibinfo {author} {\bibfnamefont {E.}~\bibnamefont {Lange}},\
  and\ \bibinfo {author} {\bibfnamefont {M.~J.}\ \bibnamefont {Rozenberg}},\
  }\href {https://doi.org/10.1103/PhysRevLett.84.5180} {\bibfield  {journal}
  {\bibinfo  {journal} {Phys. Rev. Lett.}\ }\textbf {\bibinfo {volume} {84}},\
  \bibinfo {pages} {5180} (\bibinfo {year} {2000})}\BibitemShut {NoStop}%
\bibitem [{\citenamefont {Cai}\ \emph {et~al.}(2020)\citenamefont {Cai},
  \citenamefont {Lu},\ and\ \citenamefont {Yang}}]{Cai20}%
  \BibitemOpen
  \bibfield  {author} {\bibinfo {author} {\bibfnamefont {Z.}~\bibnamefont
  {Cai}}, \bibinfo {author} {\bibfnamefont {J.}~\bibnamefont {Lu}},\ and\
  \bibinfo {author} {\bibfnamefont {S.}~\bibnamefont {Yang}},\ }\href
  {https://doi.org/10.48550/ARXIV.2006.07654} {\bibinfo {title} {Numerical
  analysis for inchworm monte carlo method: Sign problem and error growth}}
  (\bibinfo {year} {2020})\BibitemShut {NoStop}%
\bibitem [{\citenamefont {Kim}\ \emph {et~al.}(2022)\citenamefont {Kim},
  \citenamefont {Li}, \citenamefont {Eckstein},\ and\ \citenamefont
  {Werner}}]{kim_pseudoparticle_2022}%
  \BibitemOpen
  \bibfield  {author} {\bibinfo {author} {\bibfnamefont {A.~J.}\ \bibnamefont
  {Kim}}, \bibinfo {author} {\bibfnamefont {J.}~\bibnamefont {Li}}, \bibinfo
  {author} {\bibfnamefont {M.}~\bibnamefont {Eckstein}},\ and\ \bibinfo
  {author} {\bibfnamefont {P.}~\bibnamefont {Werner}},\ }\href
  {https://doi.org/10.1103/PhysRevB.106.085124} {\bibfield  {journal} {\bibinfo
   {journal} {Physical Review B}\ }\textbf {\bibinfo {volume} {106}},\ \bibinfo
  {pages} {085124} (\bibinfo {year} {2022})},\ \Eprint
  {https://arxiv.org/abs/2204.13562} {arXiv:2204.13562 [cond-mat]} \BibitemShut
  {NoStop}%
\bibitem [{\citenamefont {Prokof'ev}\ and\ \citenamefont
  {Svistunov}(2007)}]{Prokofev07}%
  \BibitemOpen
  \bibfield  {author} {\bibinfo {author} {\bibfnamefont {N.}~\bibnamefont
  {Prokof'ev}}\ and\ \bibinfo {author} {\bibfnamefont {B.}~\bibnamefont
  {Svistunov}},\ }\href {https://doi.org/10.1103/PhysRevLett.99.250201}
  {\bibfield  {journal} {\bibinfo  {journal} {Phys. Rev. Lett.}\ }\textbf
  {\bibinfo {volume} {99}},\ \bibinfo {pages} {250201} (\bibinfo {year}
  {2007})}\BibitemShut {NoStop}%
\bibitem [{\citenamefont {Cohen}\ \emph {et~al.}(2013)\citenamefont {Cohen},
  \citenamefont {Gull}, \citenamefont {Reichman}, \citenamefont {Millis},\ and\
  \citenamefont {Rabani}}]{cohen_numerically_2013}%
  \BibitemOpen
  \bibfield  {author} {\bibinfo {author} {\bibfnamefont {G.}~\bibnamefont
  {Cohen}}, \bibinfo {author} {\bibfnamefont {E.}~\bibnamefont {Gull}},
  \bibinfo {author} {\bibfnamefont {D.~R.}\ \bibnamefont {Reichman}}, \bibinfo
  {author} {\bibfnamefont {A.~J.}\ \bibnamefont {Millis}},\ and\ \bibinfo
  {author} {\bibfnamefont {E.}~\bibnamefont {Rabani}},\ }\href
  {https://doi.org/10.1103/PhysRevB.87.195108} {\bibfield  {journal} {\bibinfo
  {journal} {Physical Review B}\ }\textbf {\bibinfo {volume} {87}},\ \bibinfo
  {pages} {195108} (\bibinfo {year} {2013})}\BibitemShut {NoStop}%
\bibitem [{\citenamefont {Erpenbeck}\ \emph {et~al.}(2022)\citenamefont
  {Erpenbeck}, \citenamefont {Gull},\ and\ \citenamefont
  {Cohen}}]{Erpenbeck_Quantum_2022}%
  \BibitemOpen
  \bibfield  {author} {\bibinfo {author} {\bibfnamefont {A.}~\bibnamefont
  {Erpenbeck}}, \bibinfo {author} {\bibfnamefont {E.}~\bibnamefont {Gull}},\
  and\ \bibinfo {author} {\bibfnamefont {G.}~\bibnamefont {Cohen}},\ }\href
  {https://doi.org/10.48550/ARXIV.2207.07547} {\bibinfo {title} {Quantum monte
  carlo in the steady-state}} (\bibinfo {year} {2022})\BibitemShut {NoStop}%
\bibitem [{\citenamefont {Schir\'o}\ and\ \citenamefont
  {Fabrizio}(2009)}]{Schiro_Real_2009}%
  \BibitemOpen
  \bibfield  {author} {\bibinfo {author} {\bibfnamefont {M.}~\bibnamefont
  {Schir\'o}}\ and\ \bibinfo {author} {\bibfnamefont {M.}~\bibnamefont
  {Fabrizio}},\ }\href@noop {} {\bibfield  {journal} {\bibinfo  {journal}
  {Phys. Rev. B}\ }\textbf {\bibinfo {volume} {79}},\ \bibinfo {pages} {153302}
  (\bibinfo {year} {2009})}\BibitemShut {NoStop}%
\bibitem [{\citenamefont {Werner}\ \emph {et~al.}(2009)\citenamefont {Werner},
  \citenamefont {Oka},\ and\ \citenamefont
  {Millis}}]{Werner_Diagrammatic_2009}%
  \BibitemOpen
  \bibfield  {author} {\bibinfo {author} {\bibfnamefont {P.}~\bibnamefont
  {Werner}}, \bibinfo {author} {\bibfnamefont {T.}~\bibnamefont {Oka}},\ and\
  \bibinfo {author} {\bibfnamefont {A.~J.}\ \bibnamefont {Millis}},\
  }\href@noop {} {\bibfield  {journal} {\bibinfo  {journal} {Phys. Rev. B}\
  }\textbf {\bibinfo {volume} {79}},\ \bibinfo {pages} {035320} (\bibinfo
  {year} {2009})}\BibitemShut {NoStop}%
\end{thebibliography}%
\end{document}